\documentclass[journal, compsoc]{IEEEtran}

\usepackage{amsmath,amscd,amssymb,color,xcolor}
\usepackage{times}
\usepackage{textcomp}
\usepackage{graphicx}
\usepackage{listings}
\usepackage[ruled]{algorithm2e}
\usepackage{algorithmic}
\usepackage{paralist}
\usepackage{longtable} 
\usepackage{multirow}
\usepackage{subfigure}
\usepackage{caption}
\usepackage{mathrsfs}
\usepackage{enumitem}
\usepackage[numbers]{natbib}
\usepackage{float}
\usepackage{tcolorbox}
\usepackage{color}
\usepackage{lscape}
\usepackage{afterpage}
\usepackage{bm}
\usepackage{latexsym}
\usepackage{makecell}

\usepackage[mathscr]{eucal} 
\usepackage{tikz}
\usepackage{epstopdf}

\usepackage{adjustbox}
\usepackage{palatino}
\usepackage{url}
\usepackage{booktabs}
\usepackage{xspace}

\usetikzlibrary{automata,arrows}

\SetAlFnt{\small}
\SetAlCapFnt{\small}
\SetAlCapNameFnt{\small}
\SetAlCapHSkip{0pt}
\IncMargin{-\parindent}

\numberwithin{equation}{section} 

\allowdisplaybreaks[4]
\graphicspath{{Figures/}}
\setenumerate{fullwidth,itemindent=\parindent,listparindent=\parindent,itemsep=5pt,partopsep=0pt,parsep=0pt}

\renewcommand{\arraystretch}{2.0}

\newtheorem{definition}{Definition}

\newcommand{\zzy}[1]{\textcolor{black}{#1}}
\newcommand{\ppl}[1]{\textcolor{black}{#1}}

\newcommand{\Nskip}{\vspace{0.5em}}
\newcommand{\Sskip}{\vspace{0.5ex}}
\newcommand{\Tskip}{\vspace{0.25ex}}

\def \RP{\mathit{RP}}
\def \VR{\mathit{VR}}
\def \SFR{\mathit{SFR}}

\newcommand{\del}[1]{}

\begin{document}

\title{METTLE: a METamorphic Testing approach to assessing and validating unsupervised machine LEarning systems}

\author{Xiaoyuan~Xie,~\IEEEmembership{Member,~IEEE}, Zhiyi~Zhang, 
	Tsong~Yueh~Chen,~\IEEEmembership{Senior Member,~IEEE},
	Yang~Liu,~\IEEEmembership{Member,~IEEE}, Pak-Lok~Poon,~\IEEEmembership{Member,~IEEE}
	and Baowen~Xu,~\IEEEmembership{Senior Member,~IEEE}
	\IEEEcompsocitemizethanks{
		\IEEEcompsocthanksitem X.~Xie and Z.~Zhang are with the School of Computer Science, Wuhan University, Wuhan 430072, China.  \protect\\
		X.~Xie is the corresponding author. E-mail: xxie@whu.edu.cn.
		\IEEEcompsocthanksitem T.\,Y.~Chen is with the Department of Computer Science and Software Engineering, Swinburne University of Technology, Hawthorn, VIC 3122, Australia. 
	   \IEEEcompsocthanksitem  Y.~Liu is with the School of Computer Science and Engineering, Nanyang Technological University, Singapore 639798.
		\IEEEcompsocthanksitem P.-L.~Poon is with the School of Engineering and Technology, Central Queensland University Australia, Melbourne, VIC 3000, Australia.
	   \IEEEcompsocthanksitem  B.~Xu is with the State Key Laboratory for Novel Software Technology, Nanjing University, Nanjing 210023, China.
	}}

\maketitle

\begin{abstract}
	Unsupervised machine learning is the training of an artificial intelligence system using information that is neither classified nor labeled, with a view to modeling the underlying structure or distribution in a dataset. Since unsupervised machine learning systems are widely used in many real-world applications, \zzy{assessing the appropriateness of these systems and validating their implementations with respect to \emph{individual users' requirements and specific application scenarios\,/\,contexts}} are indisputably two important tasks. Such assessment and validation tasks, however, are fairly challenging due to the absence of a priori knowledge of the data. In view of this challenge, we develop a \textbf{MET}amorphic \textbf{T}esting approach to assessing and validating unsupervised machine \textbf{LE}arning systems, abbreviated as {\sc mettle}. 
	Our approach provides a new way to unveil the (possibly latent) characteristics of various machine learning systems, by explicitly considering the specific expectations and requirements of these systems \zzy{from individual users' perspectives}. To support {\sc mettle}, we have further formulated 11 generic metamorphic relations (MRs), covering users' generally expected characteristics that should be possessed by machine learning systems. To demonstrate the viability and effectiveness of {\sc mettle}, we have performed an experiment involving six commonly used clustering systems. 
	\zzy{Our experiment has shown that, guided by user-defined MR-based adequacy criteria, end users are able to assess, validate, and select appropriate clustering systems in accordance with their own specific needs.}
	Our investigation has also yielded insightful understanding and interpretation of the behavior of the machine learning systems from an \zzy{\emph{end-user software engineering's perspective}, rather than a designer's or implementor's perspective, who normally adopts a theoretical approach.}	
\end{abstract}
	
\begin{IEEEkeywords}
		\zzy{Unsupervised machine learning, clustering assessment, clustering validation, metamorphic testing, metamorphic relation, end-user software engineering.}
\end{IEEEkeywords}

\section{Introduction}\label{sec:introduction}

\IEEEPARstart{U}{nsupervised} machine learning requires no prior knowledge and can be widely used in a large variety of
applications such as market segmentation for targeting customers~\cite{marketing}, anomaly or fraud detection in banking~\cite{Chaudhary12areview}, 
grouping genes or proteins in biological process~\cite{gene2004},
deriving climate indices from earth science data~\cite{climate2003}, and document clustering based on content~\cite{documentclustering}. 
More recently, unsupervised machine learning has also been used by software testers in predicting software faults~\cite{fault2016}.

\zzy{This paper specifically focuses on \emph{clustering systems} (which refer to software systems that implement clustering algorithms and are intended to be used in different domains)}
Such a clustering system helps users partition a given unlabeled dataset into groups (or clusters) based on some similarity measures, so that data in the same cluster are more ``similar'' to each other than to data from different clusters. In artificial intelligence (AI) and data mining, numerous clustering systems~\cite{ Fahad2014survey,deepembedding,SAXENA2017664} have been developed and are available for public use. 
Thus, selecting the most appropriate clustering system for use is 
\zzy{an important concern from end users. (In this paper, \emph{end users}, or simply \emph{users}, refer to those people who are ``causal'' users of clustering systems. Although they have some hands-on experience on using such systems, they often do not possess a solid theoretical foundation on machine learning. These users come from different fields such as bioinformatics~\cite{rohtua2018} and nuclear engineering~\cite{nuclear2011}. Also, their main concern is the applicability of a clustering system in the 
\emph{users' specific contexts}, rather than the detailed logic of this system.)}
From a user's perspective, this selection is not 
trivial~\cite{Banerjee:2004:OEC:1014052.1014112}, \ppl{not only because end users generally do not have very solid theoretical background on machine learning, but also because the selection task} involves two complex issues as follows:

{\bf (Issue~1)} 
The correctness of the clustering results is a major concern for users.
However, when evaluating a clustering system, there is not necessarily a correct solution or "ground truth"  that users can refer to for verifying the clustering result~\cite{AW2015}.
Furthermore, not only is the correct result difficult or infeasible to find, the interpretation of correctness varies from one user to another. This is because, although data points are partitioned into clusters based on some similarity measures, the comprehension of "similarity" may vary among individual users. Given a cluster, one user may consider that the data in it are similar, yet another user may consider not.

{\bf (Issue~2)} 
Despite the importance of the correctness of the clustering result, in many cases, users would probably be more concerned if a clustering system produces an output that is appropriate or meaningful to their particular scenarios of applications. This view is supported by the following argument in~\cite{Luxburg2011art}:

\Sskip
\begin{quote}
	\emph{"... the major obstacle is the difficulty in evaluating a clustering algorithm without taking into account the context: why does the user cluster his data in the first place, and what does he want to do with the clustering afterwards? We argue that clustering should not be treated as an application-independent mathematical problem, but should always be studied in the context of its end-use."}
\end{quote}

\textbf{\emph{Regarding issue~1}}, it is well known as the \emph{oracle problem} in software testing. This problem occurs when a test oracle (or simply an oracle) does not exist. Here, an \emph{oracle} refers to a mechanism that can verify the correctness of the system output~\cite{chen2018metamorphic}.
In view of the oracle problem, users of unsupervised machine learning rely on two types of validation techniques (external and internal) to evaluate clustering systems. Basically, \emph{external} validation techniques evaluate the output clusters based on some existing benchmarks; while \emph{internal} validation techniques adopt features inherent to the data alone to validate the clustering result.

Both external and internal validation techniques suffer from some problems which affect their effectiveness and applicability. For external techniques, it is usually difficult to obtain sufficient relevant benchmark data for comparison~\cite{rendon2011internal,amigo2009comparison}. 
In most situations, the benchmarks selected for use are essentially those special cases in software verification and validation, thereby providing insufficient test adequacy, coverage, and diversity.
This issue
does not exist in internal validation techniques. However, since internal techniques mainly rely on the features associated with the dataset, their performance is easily affected by various data characteristics~%
\cite{liu2010understanding}. \emph{In addition, both external and internal techniques evaluate clustering systems mainly from the ``static'' perspective of a dataset, without considering the changeability of 
input datasets or the interrelationships among different clustering results (i.e., the "dynamic" aspect).} We argue that, in reality, users require this dynamic perspective of a clustering system  to be evaluated, because datasets may change due to various reasons. For instance, before the clustering process starts, a dataset may be pre-processed to filter
out noises and outliers for improving the reliability of the clustering result, or the data may be normalized so that different measures use the same scale for fair and reliable comparison. 

Our above argument is based on a common phenomenon that users often have some general expectations about the change in the clustering result when the dataset is changed in a particular way, for example, a better clustering result should be obtained after the noises have been filtered out from a dataset. 
To many users, evaluating this dynamic aspect (called the \emph{ripple effect of dataset change or transformation}) will give them more confidence on the performance of a clustering system than a code coverage test~\cite{chen2018metamorphic}. Despite its importance, it is unfortunate that both external and internal techniques generally do not consider the dynamic aspect of dataset transformation when testing clustering systems.

\zzy{\textbf{\emph{We now turn to issue~2}}. There has not yet been a generally accepted and systematic methodology that allows end users to effectively assess the quality and appropriateness of a clustering system for their particular applications. In traditional software testing, \emph{test adequacy} is commonly measured by code coverage criteria to unveil necessary conditions of detecting faults in the code (e.g., incorrect logic). In this regard, clustering systems are harder to assess because the logic of a machine learning model is primarily learnt from massive data. In view of this problem, a practically applicable adequacy criterion is in need to help a user assess and validate the characteristics that a clustering system should possess in a specific application scenario, so that the most appropriate system can be selected for use from this user's perspective.}
As a reminder, the characteristics that a clustering system is ``expected'' to possess may vary across different users. 
Needless to say, there is also no systematic methodology for users to \emph{validate} the appropriateness of a clustering result in their own contexts.

In view of the above two challenging issues, we propose a {\bf\small MET}amorphic {\bf T}esting approach to assessing and validating unsupervised machine {\bf\small LE}arning systems (abbreviated as {\sc mettle}). 
To alleviate Issue~1, {\sc mettle} applies the framework of metamorphic testing (MT)~\cite{chen2018metamorphic}, so that users are still able to validate a clustering system even when the oracle problem occurs. In addition, \zzy{MT is \ppl{naturally} considered to be a \ppl{candidate solution} for addressing the ripple effect of data transformation, since \ppl{MT} involves multiple inputs (or datasets) which follow a specific transformation relation}. By defining a set of \emph{metamorphic relations} (\emph{MRs}) (which capture the relations between multiple inputs (or datasets) and their corresponding outputs (or clustering results)) to be used in MT, the dynamic perspective of a clustering system can be properly assessed. 
Furthermore, the defined MRs can address Issue~2 by serving as an effective vehicle for users to specify their expected characteristics of a clustering system \ppl{in their specific application scenarios}.
\zzy{The compliance of \ppl{the clustering results across multiple system executions with these MRs} can be treated as a \ppl{practical} adequacy criterion \ppl{to help a user select the appropriate clustering system for use.}}
More details about the rationale and the procedure of {\sc mettle} will be provided in later sections.

The main contributions of this paper are summarized as follows:
\begin{itemize}
\item 
We proposed a metamorphic-testing-based approach ({\sc mettle}) to assessing and validating unsupervised machine learning systems that generally suffer from the absence of a priori knowledge of the data and a test oracle. 
Different from traditional validation methods, our approach provides a new \ppl{and lightweight machanism} to unveil the (possibly latent) characteristics of various learning systems, by explicitly considering the specific expectations and requirements of these systems from the perspective of individual users, \ppl{who do not possess a solid theoretical foundation of machine learning}. \ppl{In addition, {\sc mettle} can validate learning systems by explicitly considering the dynamic aspect of a dataset.}

\item 
We developed $11$ generic MRs to support {\sc mettle}, from users' generally expected characteristics of clustering systems. We conducted an experiment involving six commonly used clustering systems, which were assessed and compared against these $11$ MRs \zzy{through both quantitative and qualitative analysis}.   

\item
\zzy{We demonstrated a framework to help users assess clustering systems based on their own specific requirements. Guided by \ppl{an adequacy criterion (with respect to those chosen generic MRs or those MRs specifically defined by users)}, users are able to select the \ppl{appropriate unsupervised learning systems} to serve their own purposes.}

\item
\ppl{Our investigation has yielded insightful understanding and interpretation of the behaviors of some commonly used machine learning systems from a user's perspective, rather than a designer's or implementor's perspective (who normally adopts a more theoretical approach).}	

\end{itemize}

The rest of this paper is organized as follows. Section~\ref{sec:background} outlines the main concepts of clustering systems and MT\@.
Section~\ref{sec:motivation} discusses the challenges in clustering validation and the potential problems associated with dataset transformation in clustering.
Section~\ref{sec:method} describes our {\sc mettle} methodology and a list of 11 generic MRs to support {\sc mettle}\@.
Section~\ref{sec:setup} discusses our experimental setup to determine the effectiveness of {\sc mettle} in validating a set of subject clustering systems.
\zzy{
Section~\ref{sec:results} presents \ppl{an quantitative analysis of the performance of the subject clustering systems in term of their compliance with (or violation to) the 11 generic MRs, followed by an in-depth qualitative analysis on the underlying behavior patterns and plausible reasons for the violations revealed by these MRs.
Section~\ref{sec:application} illustrates how {\sc mettle} can be used as a systematic and yet easy-to-use framework (\emph{without the requirement of having sophisticated knowledge on machine learning theories}) for assessing the appropriateness of clustering systems in accordance with a user's own specific requirements and expectations.}
Section~\ref{sec:threats} \ppl{discusses some} internal and external threats to our study.
}
Section~\ref{sec:related} briefly discusses the recent related work on MT\@.
Finally, Section~\ref{sec:conclusion} concludes the paper and identifies some potentially fruitful areas for further research.  

\section{Background Concepts}
\label{sec:background}

In this section, we discuss the background concepts of clustering systems and MT\@. We also give some examples to illustrate how MT can be used as a software validation approach.

\subsection{Clustering Systems}
\label{subsec:clustervalidation}

In AI, \emph{clustering}~\cite{arbelaitz2013extensive,hennig2015handbook} 
is the task of partitioning a given unlabeled dataset into clusters based on some similarity measures, where data in the same cluster are more "similar" to each other than to data from different clusters. Thus, cluster analysis involves the discovery of the latent structure or distribution of data in a dataset. The clustering problem can be formally defined as follows:

\begin{definition}[Clustering]	
\label{def:clustering}
	Assuming that dataset $D=\{\textbf{x}_1,\textbf{x}_2,\cdots,\textbf{x}_n\}$ contains $n$ instances; each instance $\textbf{x}_i\,{=}\,(x^i_1,x^i_2,\dots,x^i_d)$ has $d$-dimensional attributes. A clustering system divides $D$ into $k$ clusters $\mathbb{C}=\{C_1,C_2,\cdots,C_k\}$ with label $L_k$ be the label for cluster $C_k$, where  $\bigcup_{i=1}^{k}C_i =D, C_i\neq \varnothing, C_i \cap C_j=\varnothing~(i\neq j, 1 \leq i,j \leq k )$. 
\end{definition}

\begin{figure} [t]
	\centering
	\includegraphics[scale=1]{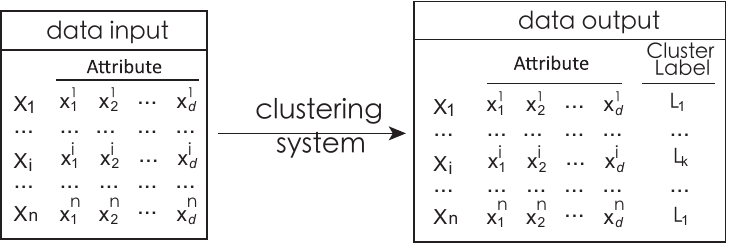}
	\caption{Clustering system.} \label{fig:clustermodel}
\end{figure}

Fig.~\ref{fig:clustermodel} describes the input and output of a clustering system.
It is well known that validating
clustering systems will encounter the oracle problem (i.e., the absence of an oracle).
For instance, it is argued in~\cite{AW2015} that:

\Nskip
\begin{quote}
\emph{"The problem is that there isn't necessarily a 'correct' or ground truth solution that we can refer to it if we want to check our answers \dots you will come to the inescapable conclusion is that there is no 'true' number of clusters (though some numbers feel better than others) [therefore a definite correct clustering result, or an oracle, does not exist], and that the same dataset is appropriately viewed at various levels of granularity depending on analysis goals."}
\end{quote}

\Nskip
\noindent
In view of the oracle problem, users of machine learning generally rely on two types (internal and external) of techniques to validate clustering systems. Both types, however, are not satisfactory because of their own limitations. \ppl{These limitations have been briefly outlined in Section~\ref{sec:introduction}, and will be further elaborated in Section~\ref{subsec:challenges}.}

\subsection{Metamorphic Testing (MT)}
\label{subsec:metamorphicmethods}

To alleviate the oracle problem, MT~\cite{chen2018metamorphic,chen1998metamorphic} has been proposed to verify and validate the ``expected'' relationships between inputs and outputs across \emph{multiple} software executions. These ``expected'' relationships are expressed as \emph{metamorphic relations} (\emph{MRs}). If the output results across multiple software executions violate an MR, then a fault is revealed. Below we give an example to illustrate the main concepts of MT and MR:

Consider a program $S$ that calculates the value of the $sin(x)$ function. It is extremely difficult to verify the correctness of the output from $S$ because an oracle is extremely difficult to compute, except that $x$ is a special value (such as $\pi$ where $sin(\pi) = 1$).
MT can help alleviate this oracle problem. Consider, for example, the mathematical property 
$sin(x) = sin(\pi - x)$. Based on this property, we can define an MR in MT: "If $y = \pi - x$, then $sin(x) = sin(y)$". With reference to this MR, $S$ is executed twice: firstly with any angle $x$ as a \emph{source test case}; and then with the angle $y$, such that $y = \pi - x$, as a \emph{follow-up test case}.
In this case, even the correct and precise value of $sin(x)$ is unknown, if the two execution results (one with input $x$ and the other with input $y$) are \emph{different} so that the above MR is violated, we can conclude that $S$ is faulty. The above example illustrates an important feature of MT\,---\, it involves \emph{multiple} software executions.

MT was initially proposed as a verification technique. 
For example, \citeauthor{murphy2008properties}~\cite{murphy2008properties} applied MT to several machine learning applications (e.g., MartiRank) and successfully revealed several defects. Different types of metamorphic properties were also categorized to provide a foundation for determining the relationships and transformations that can be used for conducting MT in machine learning applications~\cite{murphy2008properties}.
Another study has successfully demonstrated that MT can be extended to support validation of supervised machine learning software \cite{xie2011testing}. In their study,
\citeauthor{xie2011testing}~\cite{xie2011testing} presented a series of MRs (which may not be the necessary properties of the relevant algorithm) generated from the anticipated behaviors of supervised classifiers.
Violations to the MRs may indicate that the relevant classifier is unsuitable to the current application scenario, even if the algorithm is correctly implemented. 

Later, \citeauthor{zhou2016metamorphic}~\cite{zhou2016metamorphic} applied MT to validate online search services.
They adopted logical consistency relations as a measure of users' perceived quality of search services, and used this measure to validate the performance of four popular search engines such as Google and Bing. In this work~\cite{zhou2016metamorphic}, the authors compared four search engines with respect to different scenarios and factors, thereby providing users and developers with a more comprehensive understanding of how to choose a proper search engine for better searching services with clear and definite objectives.
\citeauthor{simulation2019}~\cite{simulation2019} applied MT for simulation validation, involving two prevalent simulation approaches: agent-based simulation and discret-event simulation. Guidelines were also provided for identifying MRs for both simulation approaches. Case studies~\cite{simulation2019} showed how MT can help increase users's confidence in the correctness of the simulation models.
 
MT has also been recently applied to validate a deep learning framework for automatically classifying biology cell images that involves a convolutional neural network and a massive image dataset~\cite{7961649}. This work has  demonstrated the effectiveness of MT for ensuring the quality of deep learning (especially involving massive training data). Moreover, this MT-based validation approach can be further extended for checking the quality of other deep learning applications. Other recent works~\cite{Zhou2018,Tian:2018:DAT:3180155.3180220,Zhang:2018:DGM:3238147.3238187} have also been conducted to validate autonomous driving systems where MRs were leveraged to automatically generate test cases to reflect real-world scenes.

\section{Motivation}
\label{sec:motivation}

\zzy{Recall that users of the machine learning community \ppl{often} rely on certain validation techniques (which mainly focus on the ``static'' \ppl{aspect of a dataset}) to evaluate clustering systems. Moreover, these \ppl{validiation} techniques suffer from several problems which affect their effectiveness and applicability \ppl{(e.g., unable to validate the ``dynamic'' aspect of a dataset, that is, the effect of changing the input datasets on the clustering results).} Section~\ref{subsec:challenges} below discusses in detail the limitations of most existing cluster validation techniques. Section~\ref{subsec:hiddenrisks} \ppl{then} presents some potential problems \ppl{associated} with data transformation that should be addressed when validating clustering systems.}

\subsection{Challenges in Clustering Validation}
\label{subsec:challenges}

In unsupervised machine learning, \emph{clustering} is a technique to divide a group of data samples into clusters such that data samples within the same cluster are "similar" to each other; while data samples of different clusters show ``distinct'' features from each other.
Because clustering attempts to discover hidden patterns in data with no prior knowledge, it is difficult to evaluate the correctness or quality of the clustering results (see Issues~1 and 2 in Section~\ref{sec:introduction}).

Generally speaking, there are two major types of techniques (external and internal) for validating the clustering result. Both of them, however, have their own limitations.

\Sskip
\textbf{External validation techniques.} 
The basic idea is to compare the clustering result with an external benchmark or measure, which corresponds to a pre-specified data structure. For external validity measures, there are several essential criteria to follow such as cluster homogeneity and completeness~\cite{han2011data}. Consider, for instance, the widely adopted \emph{F-measure} \cite{steinbach2000comparison}. It considers two important aspects: \emph{recall} (how many samples within a category are assigned to the same cluster) and \emph{precision} (how many samples within a cluster are in one category). It is well known that good and relevant external benchmarks are hard to obtain. This is because, in most situations, the data structure specified by the predefined class labels or other users is unknown. As a result, without prior knowledge, it is generally very expensive and difficult to obtain an appropriate external benchmark for comparing with the clustering structure generated by a clustering system.

\Sskip
\textbf{Internal validation techniques.} 
This type of techniques validates the clustering result by adopting features inherent to the dataset alone. Many internal validity indices were proposed based on two aspects: inter-cluster compactness and intra-cluster separation.
For example, one of the widely adopted indices\,---\,the \emph{silhouette coefficient}\,---\,was proposed based on the concept of distance/similarity~\cite{kaufman2009finding}.
If this coefficient (which ranges from $-1$ to $+1$) of a data sample is close to $+1$, it means that this data sample is well matched to its own cluster and poorly matched to neighboring clusters.
When compared with external techniques, on one hand, internal techniques are more practical because they can be applied without an oracle. 
On the other hand, internal techniques are less robust because they mainly rely on the features associated with the dataset, that is, data compactness and data separation. Hence, the performance of internal techniques could be easily affected by various data characteristics such as noise, density, and skewed distribution~\cite{liu2010understanding}.

\vspace{1.0ex} 
In addition to the specific limitations of external and internal validation techniques mentioned above, both types of techniques validate clustering systems mainly from a ``static'' perspective, without considering the changeability of input datasets or the interrelationships among different clustering results. 

To address the limitations of external and internal validation techniques with respect to the ``dynamic'' perspective of clustering, based on the notion of \emph{cluster stability}~\cite{doi:10.1162/089976604773717621}, various resampling techniques 
have been developed to complement the external and internal techniques. A core concept of these resampling techniques (and cluster stability) is that \emph{independent} sample sets drawn from the same underlying statistical distribution should produce similar clustering results.
Various resampling techniques~\cite{hennig2007cluster,moller2009resampling,dresen2008new} have been proposed to generate independent sample sets. 
An example of these resampling techniques is Bootstrap (a representative non-parametric resampling technique)~\cite{jain1987bootstrap}, which obtains samples by drawing a certain number of data points randomly with replacement from the original samples, and calculates a sample variance to estimate the population variance. Another example is Jittering~\cite{moller2009resampling}, which generates copies of the original sample by randomly adding noises to the dataset in order to simulate the influence of measurement errors. As a reminder, although Jittering considers noises and outliers, it does not explicitly investigate the changing trend of clusters.

To some extent, resampling techniques complement the external and internal validation techniques by comparing \emph{multiple} clustering results. 
However, it is not difficult to see from Boobstrap~\cite{jain1987bootstrap} and Jittering~ \cite{moller2009resampling} discussed above that resampling techniques do not
provide a comprehensive validation on the dynamic perspective of clustering systems,
because they mainly deal with \emph{independent} sample sets.
In reality, datasets may change constantly in various manners, involving \emph{interrelated} datasets~\cite{datasetdynamics}. 
Thus, estimating cluster stability without considering these \emph{interrelated} datasets may result in incomprehensive clustering validation.

{\bf\emph{We argue that, in most cases, users of machine learning are particularly concerned whether a clustering system produces an output that is appropriate or meaningful to their specific scenarios of applications.}} Our argument is supported by AI researchers~\cite{AW2015,Luxburg2011art}. For example, it is argued in~\cite{Luxburg2011art} that "clustering should not be treated as an application-independent mathematical problem, but should always be studied in the context of its end-use." Therefore, given a particular clustering system, one user may consider it useful, while another user may not, because of their different "expectations" or "preferences" on the clustering result. In spite of the need for catering for different users' preferences, existing clustering validation techniques (external, internal, and resampling) generally do not allow users to specify and validate their unique preferences when evaluating clustering systems (see Issue~2 in Section~\ref{sec:introduction}). Furthermore, even if we consider a particular user, it is possible that none of the existing available clustering systems fulfils all their preferences on a clustering system. If this happens, users can only choose a particular clustering system that can fulfil their preferences the best.

\ppl{It has been reported that a general, systematic, and objective assessment and validation approach for \emph{all} clustering problems does not exist~\cite{Luxburg2011art}. Although many cluster valiation methods with a range of desired characteristics have been developed, most of them are based on statistical testing and analysis. There are still other desired characteristics that existing cluster validation methods have not been addressed. In view of this problem, rather than proposing a cluster validation method which is ``generic'' enough to evaluate \emph{every} desired characteristic from \emph{all} possible users on a clustering system (which is intuitively infeasible), our strategy is to propose a ``flexible, systematic, and easy-to-use'' evaluation framework (i.e., {\sc mettle}) so that users are able to define their own sets of desired characteristics and then use these sets to validate the appropriateness of a clustering system in their specific application scenarios.}

\subsection{Potential Problems Associated with Dataset Transformation}
\label{subsec:hiddenrisks}

In reality, datasets may be changed now and then.
For example, before clustering commences, we may need to pre-process a dataset to filter out noises and outliers, in order to make the clustering result more
reliable. We may also need to normalize the data so that different measures use the same scale for the sake of comparison.
In this regard, whether data transformation may result in some unknown and undesirable ripple effect on the clustering result is a definite concern for most users. 

Often, users have some general expectations about the impact on the clustering result when the dataset is changed in a particular manner (i.e., the ``dynamic'' perspective of a dataset). Consider, for example, the filtering of noises and outliers from the dataset before clustering starts. Not only users expect the absence of the ripple effect, they also expect a better clustering result after the filtering process. Another example is that users generally expect that a clustering system is not susceptible to the input order of data. However, we observe that some clustering systems, such as $k$-means~\cite{hartigan1979algorithm}, do not meet this expectation. This is because $k$-means and some other clustering systems are, in fact, sensitive to the input order of data due to the choice of the original cluster centroid, thus even a slight offset in distance will have a noticeable effect on the clustering result.

One may argue that $k$-means is a popular clustering system, so users are likely to be aware of its above characteristic with respect to the input order of data. As a result, users will consider this issue when evaluating whether $k$-means should be used for clustering \ppl{in their own contexts}. We argue, however, as more and more new clustering systems are developed, it is practically infeasible for users to be knowledgable about the potential ripple effect of data transformation for every method (particularly the newly developed ones), so that the most appropriate one could be selected for use.

\section{Our Methodology: \large{mettle}}
\label{sec:method}

\ppl{This section introduces our approach for cluster assessment and validation. Section~\ref{features} outlines some key features and core concepts associated with {\sc mettle} from the perspective of \emph{end-user software engineering}~\cite{Segal2005, Burnett2009}. Section~\ref{definition} gives the relevant definitions used in {\sc mettle}\@. Section~\ref{subsec:MR} then presents a list of $11$ generic MRs (which are based on some common end users' expectations on a clustering system) developed to support {\sc mettle}.}

\subsection{Key Features and Core Concepts}
\label{features}

\ppl{To alleviate the challenges and potential problems mentioned in Sections~\ref{subsec:challenges} and~\ref{subsec:hiddenrisks}, we propose an MT-based methodology ({\sc mettle}) for ``users'' to assess and validate unsupervised machine learning systems. In this paper, as explained in Section~\ref{sec:introduction}, \emph{users} refer to those ``causal'' users with some hands-on experience on using clustering systems in their specific application scenarios (e.g., biomedicine, market segmentation, and document clustering), but do not possess a solid theoretical foundation on machine learning.
Thus, these users often have little interest on the internal logic of clustering systems.
Rather, they are more concerned with the applicability of these systems in their own usage contexts.
Consider, for example, users in bioinformatics consider using a clustering system to perform predictive analysis. These bioinfomaticians often have good domain knowledge about complex biological applications, but they are not experts in machine learning, or do not care much about the detailed theories in machine learning. For these users, there is a vital demand for an effective and yet easy-to-use validation method (but without the need for having sophisticated knowledge about machine learning) to help them select an appropriate clustering system for their own use.}

Some key features of {\sc mettle} are listed below:

\Sskip
\begin{itemize}
\item [(1)] 
It alleviates the oracle problem in validating clustering systems (see Issue~1 in Section~\ref{sec:introduction}).

\item [(2)]
It allows users to comprehensively assess and validate the ``dynamic'' perspectives of clustering systems related to dataset transformation. In other words, it enables users to test the impact on the clustering result when the input dataset is changed in a particular way for a given clustering system. Thus, {\sc mettle} works well with interrelated datasets.

\item [(3)]
It allows users to assess and validate their expected characteristics (expressed in the form of MRs) of a clustering system. In addition, during assessing and validating, users are able to assign weighted scores to \ppl{defined MRs} in accordance with their relative importance from the users' perspectives. \ppl{As such, an MR-based adequacy criterion, by means of a set of user's defined MRs, can be derived to help users select an appropriate clustering system for their own use} (see Issue~2 in Section~\ref{sec:introduction}).

\item [(4)]
{\sc mettle} is supported with an initial suite of 11 MRs, which are fairly generic and are expected to be applicable \zzy{across many different application scenarios and contexts.} \zzy{As a reminder, \ppl{in reality,} users may \ppl{ignore} some of these MRs that are irrelevant or inapplicable in \ppl{a specific application scenario.}}
\end{itemize}
\Sskip

Features~(1) to (3) of {\sc mettle} are made available by allowing users to define a set of MRs, with each MR captures a relation between multiple inputs (datasets) and outputs (clusters) across different clustering tasks. These user-defined MRs, together with the generic MRs in the initial suite (feature~(4) above), are assigned with weighted scores to reflect their relative importance from the user's perspective (feature~(3) above). Such "ranked" MRs thus allow users to specify their expected characteristics of a clustering system. If a clustering system generates results from multiple executions which violate an MR, it indicates that this system does not fulfill the expected characteristic corresponding to this MR\@. Thus, the set of \ppl{user-defined specific MRs and user-chosen generic MRs} essentially serves as  \zzy{a test adequacy criterion} for users to evaluate candidate clustering systems, with a view to selecting the most appropriate one for use.

\subsection{Definitions}
\label{definition}

\textbf{MR for cluster validation.}~\,%
Given a clustering system $A$ and a dataset $D$. Let $R_s\,{=}\,A(D)$ denote the clustering result.
Assume that a transformation $T$ is applied to $D$ and generates $D^T$. 	
Let $R_f\,=\,A(D^T)$ denote the new clustering result. 
An MR defines the expectation from users about the changing trend of $A$'s behaviors after transforming $D$ by $T$, that is, the expected relation $\mathbb{R}^T$ between $R_s$ and $R_f$ after $T$.

We call the original dataset $D$ and the result $R_s$ as the \emph{source input} (source sample set) and the \emph{source output} (source clustering result), respectively; call the transformed $D^T$ and the result $R_f$ as the \emph{follow-up input} (follow-up sample set) and the \emph{follow-up output} (follow-up clustering result), respectively; and call the clustering processes with $D$ and $D^T$ as the \emph{source execution} and the \emph{follow-up execution}, respectively.

\Nskip
\noindent \textbf{Output Relations.}~\,%
An MR for validation may not be a necessary property of the system under test, especially for machine learning systems. Also,
clustering results may vary due to randomness.
Thus, we will not simply check whether or not an MR holds across multiple clustering results, as normally done in MT\@. If the clustering results violate an MR, we will investigate the reason(s) for such violation.
To facilitate this, we will analyze and investigate an output relation across different clustering results in the following aspects:

\Nskip
\begin{itemize}	
	\item 
        \textbf{Changes on the returned cluster label for each sample object in the source data input $\bm{D}$.}~\,%
	For each MR, map each sample object $\textbf{x}_i^s{\in}D$ to a new object $\textbf{x}_i^f{\in}D^T$ (with changed or unchanged attribute values).
	To understand how the clustering result changes after applying the data transformation $T$, it is necessary to compare 
	the returned label for each object $\textbf{x}_i^s{\in}D$ and its corresponding object $\textbf{x}_i^f{\in}D^T$. 

       \vspace{0.5ex}	
	\item
	\textbf{Consistency between the expected label and the actual label for each newly added sample in $\bm{D^T}$.}~\,%
	Apart from mapping source data objects into their corresponding follow-up data objects, some MRs may also involve creating new objects such that users may have different expectations for the behaviors of these newly inserted objects. The newly added objects may share the same label with their neighbors, or may be assigned a new label. We will illustrate different expectations in corresponding MRs in Section~\ref{subsec:MR}.	
\end{itemize}

\Nskip
In view of the above two aspects, we propose the notion of \emph{reclustering percentage} which measures the inconsistency between a source output and its corresponding follow-up output. This notion is formally defined as follows:

\Nskip
\noindent \textbf{Reclustering percentage.}~%
Given a clustering system $A$, an MR, and a source input dataset $D\,{=}\,\{\textbf{x}_1^s,\textbf{x}_2^s, \dots, \textbf{x}_n^s\}$, by applying the data transformation $T$ to $D$ with respect to this MR, we obtain the corresponding follow-up input dataset $D^T\,{=}\,\{\textbf{x}_1^f, \textbf{x}_2^f, \dots, \textbf{x}_m^f\}$ (where $n\,{=}\,m$ if no new objects are added; $n\,{<}\,m$ if there are new objects inserted into $D^T$).
Let $d_{old}$ denote the number of cases where $\textbf{x}_i^f$ has a cluster label different from that of $\textbf{x}_i^s$ (where $1 \leqslant i \leqslant\,n$); $d_{new}$ denote the number of cases where a newly added object $\textbf{x}_j^f$ has a different cluster label than expected (where $n < j \leqslant\,m$); $|D|$ denote the size of the source input dataset; and $|D^T|$ denote the size of the follow-up input dataset.
\emph{Reclustering percentage} ($\RP$) is defined as:
\[
{\RP} = \frac{d_{old}+d_{new}}{|D^T|}
\]

\noindent 
Obviously, ${\RP}\,{=}\,0$ if no violation to MR is observed between this pair of source and follow-up executions.

It should be reminded that, in the above definition:

\Sskip
\begin{itemize}
\item We do not adopt some general similarity coefficients, such as Jaccard that calculates the intersection over union, because the $\RP$ measure we defined above serves our purpose more precisely.

\Sskip
\item
The above definition may extend beyond the necessary properties of a clustering system, because our purpose is to validate the characteristics of a clustering system instead of detecting the source code faults in its corresponding implementation.
In particular, if the clustering results generated from two related datasets do not follow the specified relation in an MR definition, a violation is said to be revealed and the characteristics of the corresponding system should be evaluated in detail to identify how and why these characteristics affect the clustering results.
\end{itemize}

\Sskip
\noindent
Also, it is not difficult to see from the above that, by configuring the transformation $T$ with various operations, the various behaviors of a clustering system can be validated.

\subsection{Generic MRs}
\label{subsec:MR}

\zzy{{\sc mettle} aims to provide an effective vehicle for end users without the need for a theoretical background of clustering to assess their expected characteristics for a clustering system, and validate the appropriateness of a clustering result in their own context.}

\zzy{To support {\sc mettle}, we developed an initial suite of 11 generic MRs. Each of these generic MRs is defined based on 
users' general expectations on the clustering result when a dataset changes in a particular way. These expectations are not gained from the theoretical background of any particular machine learning system, but from intrinsic and intuitive requirements of a clustering task. In other words, METTLE is primarily developed for users, without the need for a solid theoretical foundation of machine learning.}

These 11 generic MRs fall into six different aspects of properties of a clustering system, and are expected to be applicable across various users' perspectives. 
\zzy{\ppl{Note that these generic MRs do not cover every possible property of a clustering system that is expected by all users,} because different users may have \ppl{different sets of} expectations of a clustering system.}
\ppl{This problem, however, is not an issue in {\sc mettle} because users can, at their own will,} simply adopt any of these 11 generic MRs, and also define additional, more specific MRs for their specific scenarios of applications.

In contrast to a purely theoretical analysis on the properties of a clustering system, {\sc mettle} \ppl{takes a lightweight and more practical approach to its application.}
{\sc mettle} helps users determine the relative "usefulness" among a set of clustering systems in different specific scenarios. This, in turn, facilitates the comparison and selection of the most appropriate clustering system from the user's perspective.
Below we discuss these 11 generic MRs we developed:

\Sskip

\Sskip
\begin{itemize}
	\item [\bf (1)]\textbf{Manipulating the sample object order in the dataset.}
	Reordering sample objects is a frequently performed operation, and users often assume that this operation is trivial and, hence, does not affect the clustering result. However, this assumption is not held for some clustering systems, such as $k$-means~\cite{hartigan1979algorithm} as discussed in Section~\ref{subsec:hiddenrisks}.
	To validate whether or not this assumption is held for a clustering system, MR1.1 and MR1.2 are defined as follows:
	
	\Nskip
	\begin{quote}
		\textbf{MR1.1\,---\,Changing the object order.} 
		If we permute the order of the sample objects in the dataset, the new clustering result ($R_f$) will remain the same as the original result ($R_s$).
	\end{quote}
	
	\Nskip
	\begin{quote}
		\textbf{MR1.2\,---\,Changing the object order but keeping the same set of starting centroids.}
		If we permute the order of the sample objects in the dataset but keeping the same set of starting centroids, we will have $R_f = R_s$.
	\end{quote}
	
	\Nskip
	In MR1.2, \emph{starting centroids} are those objects that are randomly selected by a clustering system when it starts execution. Thus, by fixing the starting centroids, we can alleviate the randomness problem (i.e., the same dataset gives rise to different clustering results)
	associated with system execution. Consider, for example, $k$-means. It randomly selects $k$ objects from $D$ as the initial cluster centroids, then assigns each object
	to the cluster with the closest centroid. Clusters are then formed by recomputing cluster centroids and reassigning data objects. With respect to this property of the system, if we fix $k$ initial objects when it starts execution, followed by shuffling the other objects in $D$, it is generally expected that $R_f = R_s$, leading to MR1.2 above.
	
	\Nskip
	It should be noted that MR1.1 differs from MR1.2 in that the former may or may not involve changing the starting centroids, but the latter keeps the starting centroids unchanged.
	
	\begin{figure}[t]
		\centering
		\subfigure[MR2.1]{\label{subfig:MR2.1Description}
			\includegraphics[scale=0.7]{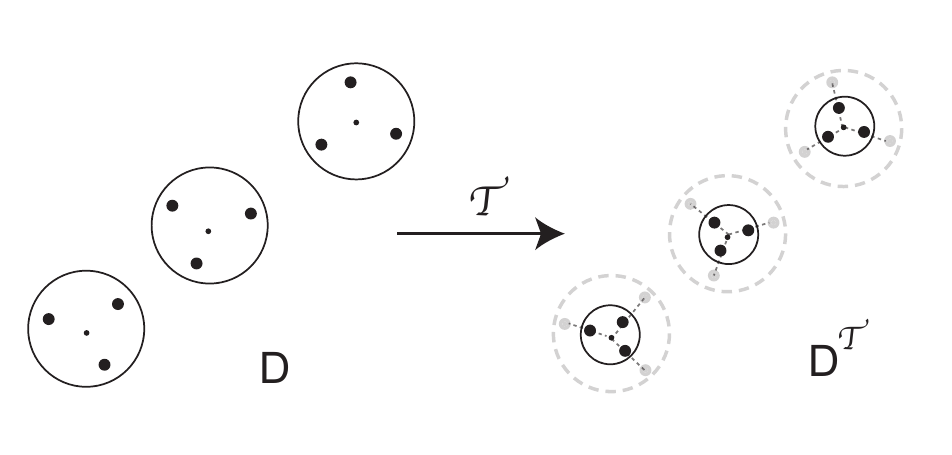}}
		\subfigure[MR2.2]{\label{subfig:MR2.2Description}
			\includegraphics[scale=0.9]{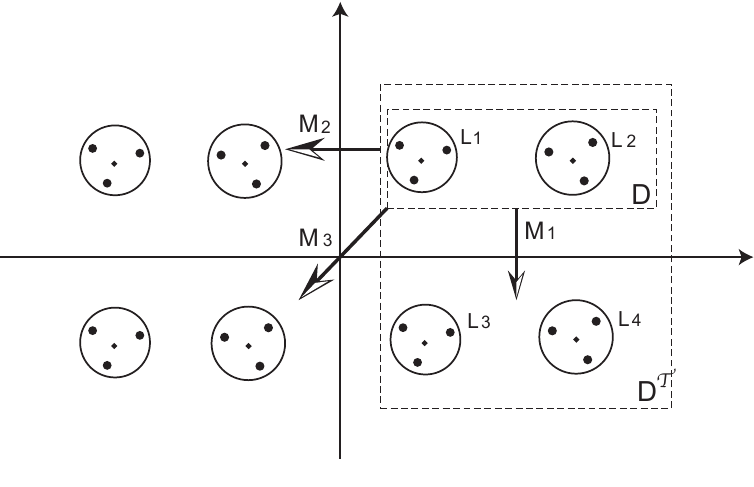}}
		\hspace{0.6in}
		\caption{Illustration on MR2.1 and MR2.2.} \label{fig:MRDescription}
	\end{figure} 
	
	\Nskip
	\item [\bf (2)]\textbf{Manipulating the distinctness among clusters in the dataset.}
	Users often expect that the distinctness among clusters will affect the clustering result. First, we consider the impact on the clustering result by shrinking some or all of the clusters towards their centroids in the dataset (see Fig.~\ref{subfig:MR2.1Description}). MR2.1 is defined accordingly as follows:
	
	\Nskip
	\begin{quote}
		\textbf{MR2.1\,---\,Shrinking one or more clusters towards their centroids.} 
		If some or all of the clusters in the dataset are shrunk towards their centroids, we will have $R_f = R_s$.
	\end{quote}
	
	\Nskip
	The rationale behind MR2.1 is obvious and needs no explanation. With respect to MR2.1, for each cluster $C_k$ in $R_s$ to be shrunk, we first identify its centroid $\textbf{m}_k$ returned by the clustering system. 
	Then, for each $\textbf{x}_i$ in $C_k$, we compute the middle point (denoted as $\textbf{x}_i^k$) from $\textbf{m}_k$ to $\textbf{x}_i$.
	$D^T$ is constructed by replacing all $\textbf{x}_i$ in $D$ with $\textbf{x}_i^k$.
	
	\Sskip
	Another aspect related to changing the distinctness among clusters is data mirroring, which is related to the following MR: 
	
	\Nskip
	\begin{quote}
		\textbf{MR2.2\,---\,Data mirroring.}
		Given an initial dataset $D$ such as its corresponding $R_s$ contains $k$ clusters in the \emph{same} quadrant of the space. 
		If we mirror all these $k$ clusters in $R_s$ to other $N$ quadrants of the space so that clusters have approximately equal distance to each other,
		then $\big((N+1)*k\big)$ clusters will be formed in $R_f$. Furthermore, the newly formed clusters in $R_f$ will include the original $k$ clusters in $R_s$. 
	\end{quote}
	
	\Nskip
	To illustrate MR2.2, let us consider a two-dimensional space in Fig.~\ref{subfig:MR2.2Description}. Suppose, after the first execution of a clustering system $A$, $R_s$ contains two clusters $L_1$ and $L_2$. We then segment the space into four quadrants, where $L_1$ and $L_2$ are in the \emph{same} quadrant. With the mirroring operation $M_1$, we mirror $L_1$ and $L_2$ (and the sample objects contained in them) in $D$ to an adjacent quadrant to create new "mirroring" clusters $L_3$ and $L_4$. A new dataset $D^T$ is created, containing the original clusters ($L_1$ and $L_2$ before mirroring) and the newly formed "mirroring" clusters ($L_3$ and $L_4$ after mirroring). We then perform two more mirroring operations ($M_2$ and $M_3$)
	 in Fig.~\ref{subfig:MR2.2Description}) similar to $M_1$ to create additional "mirroring" clusters. Finally, we perform another execution of $A$, and compare the clusters in $R_s$ and $R_f$ to see whether or not MR2.2 is violated.
	
	\Nskip
	\item [\bf (3)]\textbf{Manipulating the object density of one or more clusters in the dataset.}
	Suppose additional sample objects are added into some clusters in the dataset $D$ to increase the object densities of these clusters (see Fig.~\ref{fig:MR3.1Description}). With respect to this action, users will expect that every newly object added to a cluster $L$ (before executing the clustering system) will indeed be assigned to $L$ by the clustering system after its execution. In reality, however, not every clustering system meets such user's expectation. To validate the behavior of a clustering system with respect to the change in the object densities of clusters, we define the following MR:
	
	\Nskip
	\begin{quote}
		\textbf{MR3.1\,---\,Adding sample objects around cluster centers.} 
		If we add new sample objects to a cluster $L$ in $R_s$ so that they are closer to the centroid of $L$ than some existing objects in $L$, followed by executing the clustering system again, then: (a)~all the clusters 
		appearing in $R_s$ will also appear in $R_f$, and (b)~these newly added sample objects will also appear in $R_f$ and with $L$ as their cluster.
	\end{quote}
	
	\Nskip
	MR3.1 can be validated in a similar way as to validating MR2.1 but with some changes.
	First, similar to validating MR2.1, we create a new object $\textbf{x}_i^k$ for an existing object $\textbf{x}_i$ in a given cluster $C_k$ of $R_s$, such that $\textbf{x}_i^k$ is the middle point between $\textbf{x}_i$ and the centroid $\textbf{m}_k$. However, for validating MR3.1, we do not create $\textbf{x}_i^k$ for each $\textbf{x}_i$. 
	Rather, we randomly select $\textbf{x}_i$. 
	Secondly, the newly created $\textbf{x}_i^k$ is added as a new element, instead of replacing the original $\textbf{x}_i$ as for validating MR2.1.
	\Nskip
	
	MR3.1 can be slightly revised to create another metamorphic relation (MR3.2); the latter involves adding sample objects near the boundary of a cluster.
	
	\Nskip
	\begin{quote}
		\textbf{MR3.2\,---\,Adding sample objects near a cluster's boundary.}
		If we randomly add new sample objects on the edge of the convex hull\,%
		\footnote{~In mathematics, the \emph{convex hull} of a set $X$ of points in the Euclidean plane is the smallest convex set that contains $X$.}
		of the objects whose cluster is $L$, followed by executing the clustering system again, then: (a)~all the clusters appearing in $R_s$ will also appear in $R_f$, and (b)~these newly added objects will also appear in $R_f$ and with $L$ as their cluster.
	\end{quote}
	\Nskip
	
	\begin{figure}[t] 
		\centering
		\includegraphics[scale=0.6]{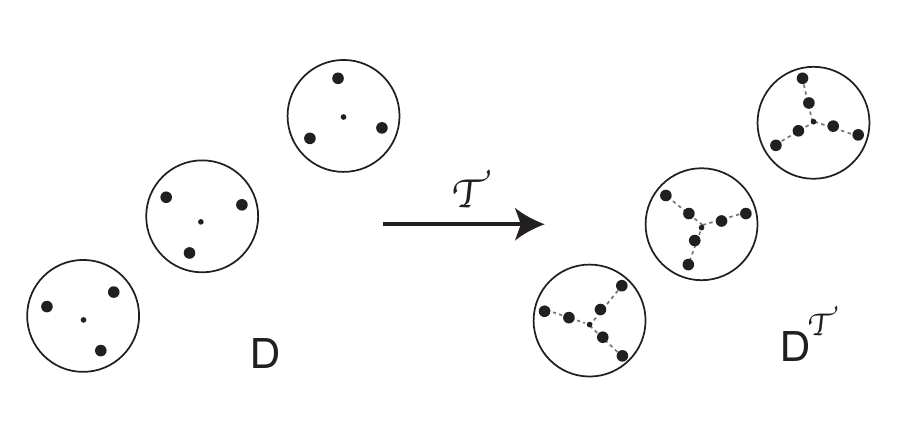}
		\caption{Illustration on MR3.1.} \label{fig:MR3.1Description}
	\end{figure}
	
	\begin{figure}[t] 
		\centering
		\includegraphics[scale=0.8]{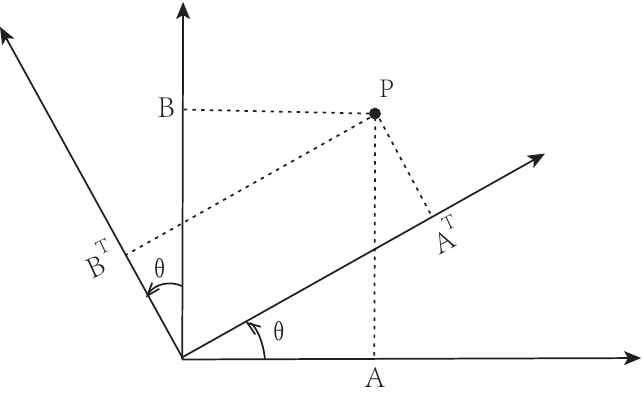}
		\caption{Illustration on MR5.1.} \label{fig:MR5.1Description}
	\end{figure}
	
	\item [\bf (4)]\textbf{Manipulating attributes.}
	Attributes in a dataset may be occasionally changed. We consider two possible types of transformation on attributes.
	First, new attributes may be added to a dataset, if they are considered representative for distinguishing sample objects. 
	In view of this addition, MR4.1 is defined as follows:
	
	\Nskip
	\begin{quote}
		\textbf{MR4.1\,---\,Adding informative attributes.} 
		We define an \emph{informative attribute} as the one whose value for each object $\textbf{x}_i\,{=}\,\{x^i_1,x^i_2,\dots,x^i_d\}$ is the corresponding returned cluster name $l_i$ in $R_s$ ($l_i$ could be any $C_k\,{\in}\,R_s$). $D^T$ is constructed by adding this new informative attribute to $D$, that is, each object $\textbf{x}_i^T$ in $D^T$ becomes $\textbf{x}_i^T\,{=}\,\{x^i_1,x^i_2,\dots,x^i_d, l_i\}$.	Then, we will have $R_f = R_s$.
	\end{quote}
	
	\Nskip
	Next, we consider the second type of data transformation.
	An attribute is generally considered as \emph{redundant}
	if it can be derived from another attribute~\cite{han2011data}.
	Redundancy is a critical issue in data integration, and its occurrence can be detected by correlation analysis. \citet{han2011data} argue that a high correlation
	generally indicates that an attribute can be considered redundant and hence to be removed. To define an MR related to redundant attributes, we adopt a widely used Pearson's product moment coefficient to measure the degree of correlation between attributes, and construct $D^T$ by removing redundant attributes (if any). Intuitively speaking, we expect removing redundant attributes will not affect the clustering result. This expectation leads to the following MR:
	
	\Nskip
	\begin{quote}	
		\textbf{MR4.2\,---\,Removing redundant attributes.} If we remove one or more redundant attributes from the dataset $D$ and then execute the clustering system again, we will have $R_f = R_s$.
	\end{quote}
	
	\Nskip
	\item [\bf (5)]\textbf{Manipulating the coordinate system.} Several ways exist for manipulating the coordinate system such as rotation, scaling, and translation. These ways of changing the coordinate system shall not affect the spatial distribution of sample objects, thereby leading to the next two MRs:
	
	\Nskip
	\begin{quote}
		\textbf{MR5.1\,---\,Rotating the coordinate system.}
		Suppose the original coordinates are $(A, B)$. We perform a transformation $T$ by rotating the coordinate system by a random degree $\theta$ (where $\theta \in [0^\circ,90^\circ]$) anticlockwise. After performing $T$, we get the new coordinates $(A^T, B^T)$.
		The same set of clusters will appear in both $R_s$ and $R_f$.
	\end{quote}
	
	\Nskip
	Fig.~\ref{fig:MR5.1Description} depicts the transformation $T$. The formula below can be used to transform the existing coordinates in $D$ into the corresponding new coordinates in $D^T$:
	
	\begin{equation*}
	\begin{pmatrix}
	\begin{smallmatrix}
	A^T & B^T
	\end{smallmatrix}
	\end{pmatrix}  
	= 
	\begin{pmatrix}
	\begin{smallmatrix}
	A & B
	\end{smallmatrix}
	\end{pmatrix} 
	\begin{pmatrix}
	\begin{smallmatrix}
	cos\theta~ & -sin\theta \\
	sin\theta~ & ~cos\theta
	\end{smallmatrix}
	\end{pmatrix}
	\end{equation*}
	
	\Sskip
	A scaling transformation changes the sizes of clusters. Scaling is performed by multiplying the original coordinates of objects with a scaling factor.
	
	\Nskip
	\begin{quote}
		\textbf{MR5.2\,---\,Scaling the coordinate system.} Suppose the original coordinates are $(A, B)$; the scaling factors for the two axes are $S_a$ and $S_b$, respectively; and the new coordinates after scaling are $(A^T, B^T)$ (the mathematical representation of this scaling transformation is shown in the formula below).
		\zzy{When $S_a {=} S_b$, we will have $R_f = R_s$.}
	\end{quote}
	
	\begin{equation*}
	\begin{pmatrix} 
	\begin{smallmatrix}
	A^T & B^T
	\end{smallmatrix} 
	\end{pmatrix}
	=
	\begin{pmatrix} 
	\begin{smallmatrix}
	A & B 
	\end{smallmatrix} 
	\end{pmatrix}
	\begin{pmatrix} 
	\begin{smallmatrix}
	S_a~ & 0 \\
	0~ & S_b
	\end{smallmatrix} 
	\end{pmatrix}
	\end{equation*}
	
	\Sskip	
	\item [\bf (6)]\textbf{Manipulating outliers.}
	An \emph{outlier} is a data object that acts quite different from the rest of the objects, as if it were generated by a different mechanism \cite{han2011data}. 
	It is generally expected that a clustering system will handle outliers by either filtering them or assigning new cluster labels to them. In our study, we mainly focus on \emph{global outliers}, which do not follow the same distribution as other sample objects and significantly deviate from the rest of the dataset~\cite{han2011data}.
	
	\Nskip	
	\begin{quote}
		\textbf{MR6\,---\,Inserting outliers.}  
		To generate $D^T$, we add a sample object $X_o$ to the dataset $D$  so that the distance from $X_o$ to any cluster
		is much larger than the average distance between clusters (in order to make $X_o$ not associated with any predefined clusters in $D$). After this operation, the following properties must be met:
		(a)~every object (except $X_o$) has the same cluster label in both $D$ and $D^T$, and
		(b)~$X_o$ does not occur in $R_f$, or if $X_o$ occurs in $R_f$ then $X_o$ has a new cluster label which is not associated with all the other objects.
	\end{quote}
	
\end{itemize}


\section{Experimental Setup}
\label{sec:setup}

\ppl{This section outlines the setup of our experiment, which follows the guidelines by~\citeauthor{wohlin2012}~\cite{wohlin2012} as far as possible. In what follows, we first define the main objective and research questions of our experiment. This is followed by discussing the subject clustering systems used in the experiment. Thereafter, we discuss the detailed experimental design, including environment configuration, experimental procedures, dataset preparation, and parameter setting.} 


A few properties \ppl{corresponding to some of the generic MRs discussed in Section~\ref{subsec:MR}} were \emph{individually} investigated in some previous studies. For example, it has been reported in~\cite{omran2007overview} that the performance of $k$-means depends on the initial dataset conditions. More specifically, some initial dataset conditions may cause $k$-means to produce suboptimal clustering results.
As another example, density-based clustering systems are found to be generally efficient at separating noises and outliers~\cite{ester1996density}. 
However, few work has been done to provide a systematic, practical, and lightweight approach for validating a set of clustering systems with reference to various properties (defined from the user's perspective) in a comprehensive and holistic manner.

\subsection{Research Objective and Questions}
\label{subsec:RQs}

The main objective of our experiment is to demonstrate, by means of quantitative and qualitative analyses, the feasibility and practicality of {\sc mettle} for assessing and validating clustering systems with respect to a set of system characteristics as defined by users.
In this paper, we do not intend to conduct a comparative experiment with other "traditional" cluster validation techniques. This is because most of these techniques take a statistical perspective while {\sc mettle} focuses on the users' perspective; this difference in perspective renders a comparison meaningless.

In view of the above research objective, the following two research questions have been set:

\Sskip 
\begin{itemize}
	\item {\bf RQ1:} What is the performance of each subject clustering system with respect to the 11 generic MRs?
	\Sskip 
	\item {\bf RQ2:} What are the underlying behaviors of the subject clustering systems that cause violations to the relevant MRs (if any)?
\end{itemize}

\subsection{Subject Clustering Systems} 
\label{subsec:algorithms}

\zzy{Our experiment involved six popular clustering systems obtained from the open source software Weka (version 3.6.6)~\cite{witten2017data}. These six subject systems fall into three categories: prototype-based, hierarchy-based, and density-based. }

\subsubsection{Prototype-based Systems}
\label{subsec:prototype}
	Given a dataset $D=\{\textbf{x}_1, \textbf{x}_2,\cdots, \textbf{x}_n\}$ that contains $n$ instances; each instance has $d$ attributes. The main task of prototype-based systems is to find a number of representative data objects (known as \emph{prototypes}) in the data space. More specifically, an initial partition of data is built first, then a prototype-based system will minimize a given criterion by iteratively relocating data points among clusters. In this category, we specifically considered the following three methods:

\noindent
{\bf\emph k}\textbf{-means (KM).} Let ${\textbf{m}}^{(t)}$ denote the cluster centroid of each cluster, where $t$ is the number of iterations. In essence, KM~\cite{hartigan1979algorithm} involves the following major steps:
		\begin{enumerate}
			\item [(1)] Randomly choose $k$ data points as the initial cluster centroids.
			\item [(2)] Assign each data point to the nearest centroid, using the following formula (in which $\Arrowvert \Arrowvert$ means the L2 norm):

			\begin{equation*}
			{C_i}^{(t)}=\{\textbf{x}_p: \Arrowvert \textbf{x}_p -{\textbf{m}_i}^{(t)} \Arrowvert \leq \Arrowvert \textbf{x}_p -{\textbf{m}_j}^{(t)} \Arrowvert \}
			\end{equation*}

The above formula follows the notation in Definition~\ref{def:clustering} in Section~\ref{subsec:clustervalidation}, where ${C_i}^{(t)}$ denotes the $i$th cluster in the $t$th iteration, and $C_i$ denotes a set of points whose label is the current cluster.

			\item [(3)] Recalculate the centroid of each cluster, the new centroid is as follows:

			\begin{equation*}
			{\textbf{m}_i}^{(t+1)}={\frac{1}{\lvert {C_i}^{(t)}\rvert}}\sum_{
				\textbf{x}_j\in {C_i}^{(t)}} \textbf{x}_j 
			\end{equation*}

			where $\textbf{m}_i$ is the centroid of cluster $C_i$, and ${\textbf{m}_i}^{(t+1)}$ is the new centroid.

			\item [(4)] Repeat steps~(2) and (3) above until there is no further change in clusters or the predefined maximum number of iterations is reached.
		\end{enumerate}

\noindent		
{\bf\emph x}\textbf{-means (XM).} This system addresses two weaknesses of KM: (a)~poor calculation ability, and (b)~the need for foreknowing the value of $k$ and the local minima~\cite{pelleg2000x}.
Unlike KM, XM only needs users to specify a range of $k$ values so that XM can arrive at an optimal cluster number. The major steps of XM are as follows:
		\begin{enumerate}
			\item [(1)] Run conventional $k$-means, where $k$ equals to the lower bound $k_{min}$ of the specified range.
			\item [(2)] Split some centroids into two by calculating the value of the Bayesian Information Criterion (BIC) \cite{Konishi2008}.
			\item [(3)] Repeat (1) and (2) until $k\,>\,k_{max}$, where $k_{max}$ is the upper bound of the specified range.
		\end{enumerate}

\noindent		
\textbf{Expectation-Maximization (EM).} This system aims at finding the maximum likelihood of parameters in a statistical model~\cite{redner1984mixture}. EM consists of the following major steps:
		\begin{enumerate}
			\item [(1)] Initialize the distribution parameter $\theta$.
			\item [(2)] E-step: Calculate the expected value of the unobserved variable $z^{(i)}$ with respect to the current estimate of the parameter $\theta$, thereby indicating the class to which the data object $i$ belongs:

			\begin{equation*}
			Q_i(z^{(i)})=p(z^{(i)} | x^{(i)}; \theta)
			\end{equation*}
			\item [(3)] M-step: Find the parameter that maximizes the log likelihood function using the following formula:
			\begin{equation*}
			\theta={\arg \max}_{\theta} {\sum_i \sum_{z^{(i)}} Q_i(z^{(i)})log \frac{p(x^{(i)},z^{(i)};\theta)}{Q_i(z^{(i)})}}
			\end{equation*}
		\end{enumerate}

\Sskip
\subsubsection{Hierachy-based Systems}
\begin{figure}[t]
		\centering
		\subfigure[Agglomerative nesting]{ \label{subfig: AN}
			\includegraphics[scale=0.8]{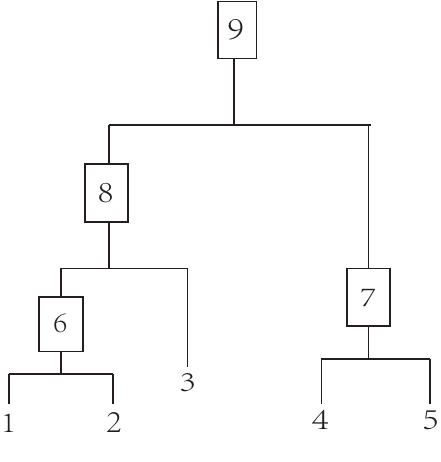}}
		\hspace{0.5in}
		\subfigure[Farthest-first traversal]{ \label{subfig: FF}
			\includegraphics[scale=0.7]{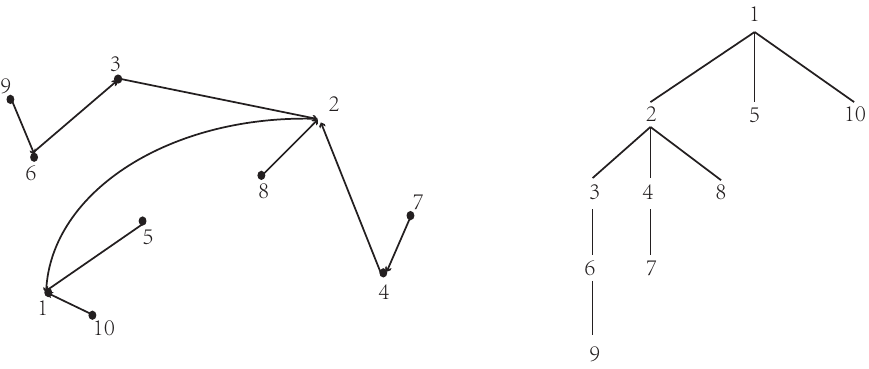}}
		\caption{Examples of clustering results generated by hierarchy-based clustering systems.}\label{fig:hierarchymethods}
	\end{figure}

	This category of systems aims at building a hierarchy of clusters by merging or splitting data partitions, and the results are usually presented as a dendrogram. Fig.~\ref{fig:hierarchymethods} shows the resulting clusters generated by two popular hierarchy-based methods: agglomerative nesting and farthest-first traversal.
	
\noindent
\Sskip
\textbf{Agglomerative nesting (AN).} 
This system adopts a bottom-up approach, where each data object is initially considered as a cluster in its own and then pairs of clusters are successively merged where appropriate. 
The clustering process has the following steps:
    \begin{enumerate}
    	\item [(1)] Assign each data point to a single cluster.
    	\item [(2)] Evaluate the pairwise distance between clusters by a distance metric (e.g., the Euclidean distance) and a linkage criterion. 	   	
    	\item [(3)] Merge the closest two clusters into one cluster according to the calculated distance.
    	\item [(4)] Repeat steps (2) and (3) above until all relevant clusters have been merged into a single cluster that contains all data points. The clustering result of AN is typically visualized as a dendrogram as shown in Figure~\ref{subfig: AN}.
	 \end{enumerate}

The linkage criterion (denoted as $\mathit{linkType}$) used in step (2) determines the distance between sets of observations as a function of the pairwise distances between these observations. Some commonly used criteria for $\mathit{linkType}$ are single-linkage, complete-linkage, and average-linkage.	
Single-linkage could lead to a bad behavior known as ``chaining'', while complete-linkage, being an opposite extreme of single-linkage, suffers from the problem of "crowding"~\cite{dasgupta2005performance}.
Take average-linkage as an example, the distance between two clusters is calculated as follows:
\begin{equation*}
d(C_i, C_j)=\frac{1}{|C_i||C_j|}\sum_{x_p \in C_i} \sum_{x_q \in C_j} d(\textbf{x}_p,\textbf{x}_q)
\end{equation*}				

AN does not require a pre-specified number of clusters (i.e., $k$). However, the dendrogram should be cut at some point if we want a partition of disjoint clusters.  Some criteria can be used to determine the cut point such as similarity level, or just a specific $k$,  which is preferred in our approach.

\Sskip
\noindent		
\textbf{Farthest-first traversal (FF).} It consists of the following three main steps~\cite{dasgupta2005performance}:

		\begin{enumerate}
			\item [(1)] Randomly pick a point from $n$ data points as a starting point and label it as $1$.
			\item [(2)] Number the remaining points using FF:
			For $i\,{=}\,2,\,3,\,\cdots,\,n$, find the unlabeled point furthest from the set \{$1,\,2,\,\cdots,\,i-1$\} and label it as $i$ (using the standard notion of distance from a point to a set: $d(x,S)=min_{y\in S} d(x,y)$). 
			For point $i$, let: $\pi (i) = argmin_{j<i} d(i,j)$ be its $parent$, and $R_i = d(i,\pi (i))$ be its distance to $\pi (i)$. A tree $T^{\pi}$ is then constructed on nodes \{$1,\,2,\,\cdots,\,n$\}, rooted at $1$ and with an edge between each point $i$ and its parent $\pi (i)$. An example is shown in Figure~\ref{subfig: FF}.
			\item [(3)] Obtaining the ordering of points (i.e., $T^{\pi}$) from step (2). The first $k$ points are regarded as $k$ cluster centers, where the remaining points in $T^{\pi}$ are assigned to their closest centers.
		\end{enumerate}
\Sskip

\subsubsection{Density-based Systems}
Many clustering systems are distance-based, thereby exhibiting the limitation on discovering non-convex clusters. On the other hand, density-based clustering systems (implemented under a data connectivity criterion) can efficiently identify clusters of arbitrary shape. 
We found two density-based clustering systems in Weka 3.6.6: DS and OPTICS (Ordering Points To Identify the Clustering Structure). Since OPTICS does not deliver the clustering result explicitly, we only chose DS in our experiment.

\Sskip
\noindent
\textbf{Density-based spatial clustering of applications with noise (DS).} Given a dataset with $n$ points in a space, DS groups data points in high density areas. Data points are labeled one of the following three types:

\Sskip
\begin{itemize}		
\item \emph{Core points:} A point $\textbf{m}$ is labeled as  \emph{core} if there exist at least a minimum number of points ($\mathit{minPts}$) that are within the specific distance $\mathit{eps}$ of $\textbf{m}$.
Also, these points are said to be directly reachable from $\textbf{m}$.
The number of points whose distances from $\textbf{m}$ are smaller than $\textit{eps}$ is called \emph{density}.
\Tskip		
\item \emph{Density-reachable points:} A point $\textbf{n}$ is said to be \emph{density-reachable} from $\textbf{m}$ if there exists a path of points $p{=}\{t_1,\,t_2,\,\ldots,\,t_k\}$, where $t_1 = m$, $t_k = n$, and for any $t_i$ in $p$, $t_{i+1}$ is directly reachable from $t_i$.
\Tskip		
\item \emph{Noisy points:} A point is marked as \emph{noise} if it is unreachable from any other points.
\end{itemize}	
	
\Sskip
\noindent
DS involves the following three main steps:
		\begin{enumerate}
			\item [(1)] Randomly select an unvisited point from the dataset.
			\item [(2)] If the selected point is a core point, then label all its density-reachable points as one cluster. Otherwise, exit.
			\item [(3)] Repeat steps~(1) and~(2) above until all points have been visited.
		\end{enumerate}

\subsection{Experimental Design}

	\subsubsection{Environment Configuration}
	\label{environment}
	
	The experimental environment \ppl{was configured as follows. Hardware environment: Intel(R) Core(TM) $i7$ CPU with 8 GB memory. Operating system: Windows 10 X64. Software development platform: Java.}

\Sskip 
	\subsubsection{Experimental Procedures}
	\label{process}
	
    Our experiment involved two main steps as follows:
		
	{\bf Step~1:} We evaluated the performance of each subject clustering system with respect to the 11 generic MRs as discussed in
	Section~\ref{subsec:MR} ({\bf RQ1}). In particular, for each system, we measured the extent of violations to these generic MRs. In general, the fewer the violations an MR reveals, the better a clustering system fits the requirement (which is expressed in that MR) of a user. To measure the extent of violation, we used two metrics: (a)~\emph{Violation Rate (VR)}\,---\,it is the ratio of the number of violated trials to the total number of trials; (b)~\emph{Reclustering Percentage (RP)}\,---\,it is the ratio of the number of objects being reassigned to the total number of objects within the follow-up dataset (previously defined in Section~\ref{definition}). We used the mean value of \emph{RP} across all trials (with different dataset per trial) to measure the extent of a clustering system that violates an MR\@. To reduce the effect of irrelevant factors on the measurement, we followed the ``blocking'' design principle~\cite{wohlin2012} in our experiment.
	
	{\bf Step~2:} For any violation to an MR, we investigated and analyzed the underlying behaviors of subject clustering systems that cause such violation, and analyzed the plausible reasons  ({\bf RQ2}). Here we carefully examined the clustering results of both source and follow-up executions, with a view to identifying their corresponding clustering patterns. This facilitated us (and users) to better understand the relevant anomalous execution behaviors of a clustering system. The investigation result was then used to develop a list of strengths and weaknesses (with respect to the 11 generic MRs) for the six subject clustering systems.

\subsubsection{Dataset Preparation and Parameter Setting}
\label{subsec:datapreparation}

For the rest of the paper, we call the dataset used for the \emph{first} execution of a clustering system the \emph{source dataset}, and the dataset (that has been changed according to a particular MR) used for the \emph{second} system execution the \emph{follow-up dataset}.

After selecting the subject clustering systems, we prepared a source dataset with clustered samples using the function $\mathit{make\_blobs}$ in Scikit-learn \cite{pedregosa2011scikit}. This function generates isotropic Gaussian blobs for clustering, that is, each cluster is a Gaussian distribution around a center point to ensure that the whole dataset is well clustered.

Let $\mathit{cluster\_std}$ denote the standard deviation of the clusters, $\mathit{centers}$ denote the number of centers to generate (default $=$ 3), $\mathit{n\_\,features}$ denote the number of features for each sample, and $\mathit{n\_sample}$ denote the total number of points equally divided among clusters. We set $\mathit{cluster\_std}$, $\mathit{centers}$, and $\mathit{n\_\,features}$ to $0.5$, $3$, and $2$, respectively. We also set $\mathit{n\_samples}$ to a valid range of [50, 200] because the larger the dataset was, the more likely violations to MRs were revealed. 

Note that there were some special cases with specific arrangements. For MR2.2, only two well-separated clusters were generated with $\mathit{cluster\_std} = 0.5$, and were mapped to the adjacent quadrant. As a result, altogether four distinctive clusters were generated in the follow-up dataset. For MR4.2, we generated an extra correlated attribute ($A'$) with a particular Pearson correlation coefficient: each sample object is three-dimensional and is denoted as $(A,\,B,\,A')$, and $\mathit{Pearson}(A,\,A')\,=\,0.8$.~\,Let $\textbf{x}_i^s\,{=}\,(A,\,B,\,A')$ be a source sample, and $\textbf{x}_i^f\,{=}\,(A,\,B)$ be its follow-up sample. Note that, given a follow-up sample, the correlated attribute $A'$ was removed from it to form its corresponding source sample. \zzy{In MR5.2, $S_a$ was randomly selected from \ppl{the} range $[0.2, 5]$ and $S_b$ \ppl{was} set to the same value \ppl{as} $S_a$.}

Based on each identified MR, follow-up datasets were derived from the corresponding source datasets. We ensured the object orders in the source datasets and follow-up datasets were properly aligned (except for MR1.1 and MR1.2 since both MRs involved changing the object orders).
Because our experiment did not focus on the effect of the input parameters of the clustering systems, we fixed the parameters in each batch of experiments: 
(a)~Euclidean distance was taken as the distance function for systems that require a distance metric; 
(b)~$\mathit{linkType}$ was set to ``AVERAGE'' (i.e., average-linkage criteria); (c)~$\mathit{eps}$ and $\mathit{minPts}$ were set to 0.1 and 8, respectively, for DS; and (d)~$\mathit{random~seed}$, which is used for random number generation in Weka implementation, was also fixed across multiple executions of each subject system involving the source datasets and their corresponding follow-up datasets, in order to ensure the clustering results were reproducible and the initial conditions were properly aligned.

As explained in Section~\ref{subsec:algorithms}, EM and DS do not need a prespecified cluster number.
 For the other four subject clustering systems, we set the parameter $k$ as follows:

\Sskip
\begin{itemize}
\item KM: Since $\mathit{centers}$ was set to $3$ (i.e., three source clusters), $k$ was also set to 3 for all MRs except MR2.2.
\Sskip
\item XM: The permissible range was set to $[k-1,\,k]$. $k$ was the actual number of clusters in a dataset and was set to $3$ except MR2.2. 
\Sskip
\item AN and FF: $k$ was set to $3$ for all MRs except MR2.2 and MR6.
\end{itemize}

\section{Experimental Results}
\label{sec:results}

\ppl{This section presents our experimental results for the two research questions RQ1 and RQ2. Section~\ref{sec:overallresults} addresses RQ1 by providing and discussing the relevant quantitative statistics. Section~\ref{sec:discussion} addresses RQ2 by providing an in-depth qualitative analysis, framed by a set of clustering patterns that we observed in the experiment. In addition, Section~\ref{subsec:summary} summarizes our observations and provides further analysis and interpretation of the results.}

\vspace{-0.5em}
\subsection{Performance of Subject Clustering Systems (RQ1)}
\label{sec:overallresults}

With respect to each of the six subject clustering systems,
we conducted $100$ trials \zzy{(with different datasets in each trial)} for each of the 11 generic MRs defined in Section~\ref{subsec:MR}.
When validating a clustering system against an MR, an experimental trial was said to cause a ``violation'', if its corresponding reclustering percentage ($\RP$) was greater than zero (see Section~\ref{definition} for the details). This result indicated that there was at least one sample reclustered "unexpectedly" in the current trial. Also, a method was said to violate an MR if there was one or more violations in all the $100$ experimental trials. 

\begin{figure}[t]
	\centering
	\includegraphics[scale=0.1]{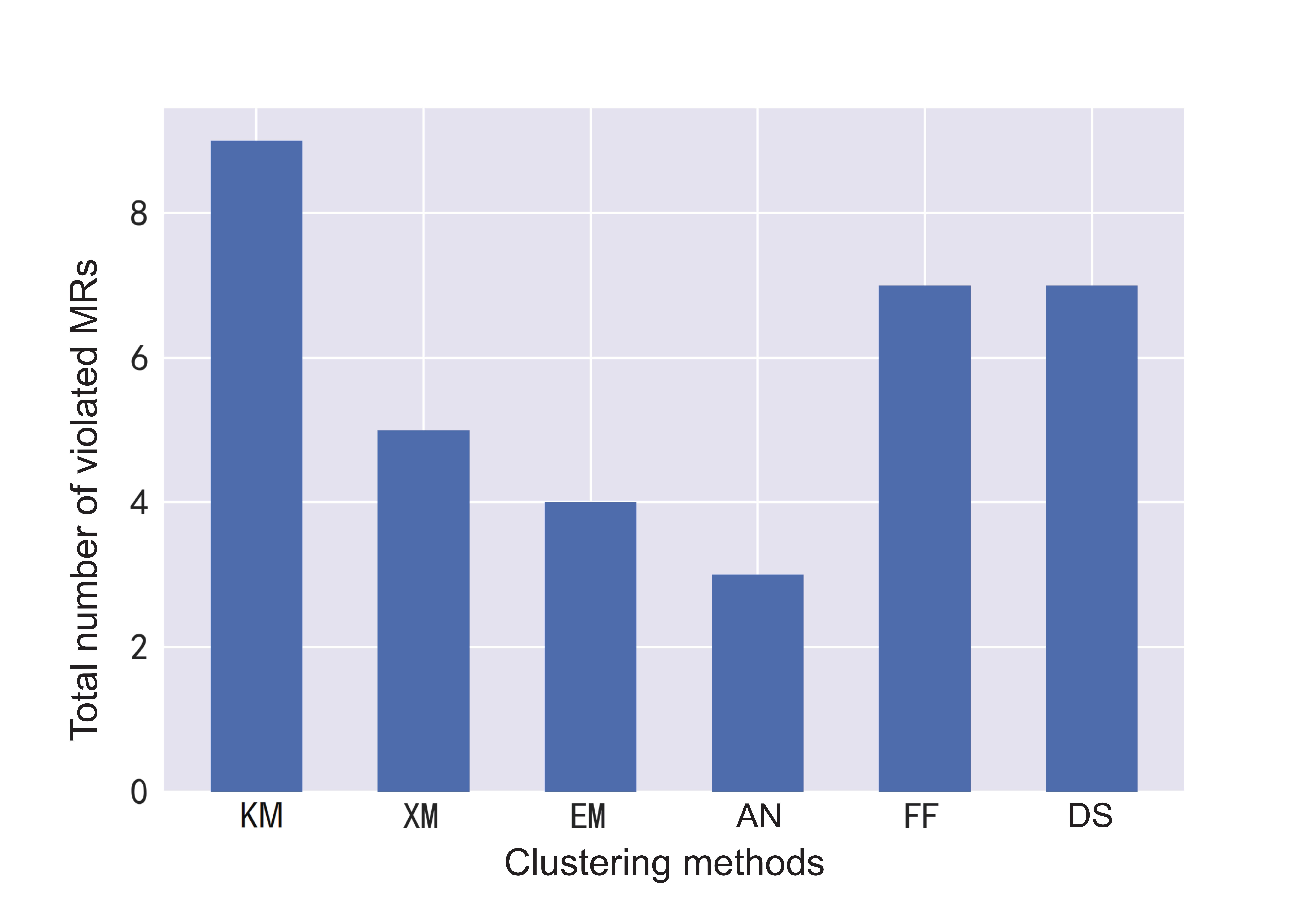}
	\hspace{0.02in}
	\caption{Total number of violated MRs of each clustering system. 
	}\label{fig:KRonalgorithm}
\end{figure}

Fig.~\ref{fig:KRonalgorithm} summarizes the total number of violated MRs of each system.
The figure shows that KM had the worst performance in that it violated nine MRs. It was followed by FF and DS\,---\,each of them violated seven MRs. XM and EM violated five and four MRs, respectively. AN performed the best because it had the smallest number of violations ($= 3$). 
Recall that every generic MR defined in Section~\ref{subsec:MR} involves data transformation in a certain way. Thus, in general, 
Fig.~\ref{fig:KRonalgorithm} indicates that KM is the most sensitive to data transformation, whereas AN is least sensitive.

Furthermore, we noted that even if two systems both violated the same MR, the chance of revealing an violation could be quite diverse. Therefore, we define the concept "violation rate" to facilitate a deeper analysis. Basically, 
\emph{violation rate} ($\VR$) is defined as the number of violation trials to all the 100 trials.
Table~\ref{tb:VR} shows the values of $\VR$ for all methods with respect to each generic MR\@.
\begin{table}[t]
	\renewcommand\arraystretch{1}
	\centering
	\caption{$\VR$ Values for Subject Clustering Systems with respect to Generic MRs} \label{tb:VR}
	\scalebox{0.8}{
		\begin{tabular}{cccccccc} 
			\toprule
			\zzy{Type of MRs} & MR & \multicolumn{3}{c}{Prototype-based} &\multicolumn{2}{c}{Hierarchy-based} & Density-based \\
			\cmidrule{3-5}  \cmidrule{6-8}
			& & KM & XM &EM &AN &FF & DS  \\
			\midrule
			\multirow{2}{0cm}{1} &
			1.1  & 5\% & 8\% & 0 & 0 & 90\% & 8\%  \\
			& 1.2 & 0 & 0 & N/A & N/A & 0 & N/A \\
			\hline
			\multirow{2}{0cm}{2} &
			2.1  & 26\% & 0 & 0 & N/A & 0 & N/A     \\
			& 2.2  & 35\% & 0 & 0 & 0 & 62\% & 100\%   \\
			\hline
			\multirow{2}{0cm}{3} &
			3.1 & 7\% & 9\% & 5\% & 14\% & 95\% & 28\% \\
			& 3.2 & 6\% & 10\% & 11\% & 15\% & 85\% & 21\%  \\
			\hline
			\multirow{2}{0cm}{4} &
			4.1  & 36\% & 0 & 0 & 0 & 8\% & 0   \\
			& 4.2  & 17\% & 0 & 0 & 0 & 0 & 7\%    \\
			\hline
			\multirow{2}{0cm}{5} &
			5.1 & 57\% & 54\% & 9\% & 47\% & 91\% & 61\% \\
			& 5.2 & 0 & 0 & 0 & 0 & 0 & 0 \\
			\hline
			\multirow{1}{0cm}{6} &
			6  & 11\% & 12\% & 39\% & 0 & 92\% & 10\% \\			
			\bottomrule
		\end{tabular}}	
		
		\vspace{-0.1cm}	
	\end{table}

Consider, for example, in this table, $\VR = 26\%$ for KM with respect to MR2.1\@. It indicates that, among the 100 experimental trials, 26 of them had their $\RP$ values greater then zero. Consider another example. $\VR = 0$ for XM with respect to MR2.1, indicating that none of the 100 experimental trials violated MR2.1. As a reminder, if "N/A'' is indicated for a particular MR in Table~\ref{tb:VR}, it means that this MR is not applicable for the relevant system(s). For instance, MR2.1 requires cluster centroids to be returned by a system. Since AN and DS do not return any cluster centroid, so their corresponding $\VR$ values are labeled as "N/A ''. 

\textbf{Zero violation.} 
Several systems had zero $\VR$ values for some MRs in Table~\ref{tb:VR}.
These zero-violation cases not only indicate a high adaptability and robustness of the corresponding systems with respect to particular types of data transformation, but they also imply that the relevant MRs may be \emph{necessary} properties of these systems and, hence, can be used for verification~\cite{xie2011testing}.
Consider, for example, the zero $\VR$ value of AN with respect to MR1.1\@. We can indeed prove that MR1.1 is a necessary property of AN\@. In this system, each data point is first considered a single cluster. Then, AN calculates the distances between clusters and incrementally merges two closest clusters. It is obvious that the distance calculation and the way of merging clusters are unrelated to the order of the data in the dataset.
In addition, Table~\ref{tb:VR} shows that no violation to MR5.2 occurred across all the six subject systems. Thus, it can be argued that MR5.2 can be considered a necessary property of the six systems.
Since this paper mainly focuses on validation rather than verification, therefore the formal proofs and analyses for zero-violation cases are excluded from this paper. Note that this paper mainly focuses on the non-zero violation cases.

\textbf{Non-zero violation.} Table~\ref{tb:VR} shows that the the non-zero $\VR$ values spread across a wide range from 5\% to 100\%. 
Intuitively speaking, with respect to an MR: (a) a high $\VR$ value indicates that a system is very sensitive to the type of data transformation corresponding to this MR, and the clustering result is likely to vary unexpectedly; and (b) a low $\VR$ value indicates that a system is relatively robust to the corresponding data transformation, and violations to this MR occur sporadically among all the experimental trials. 
Consider, for example, the values of $\VR$ of KM (5\%) and FF (90\%) with respect to MR1.1\@. The result indicates that KM violated MR1.1 in only five trials out of 100, while FF violated as many as 90 trials out of 100. Thus, the result shows that FF is far more sensitive to the type of data transformation corresponding to MR1.1 (i.e., changing the object order) than KM\@.

By examining how a system reclusters transformed data samples in each violated case, we observed that different cases had different levels of inconsistency as measured by $\RP$. 
In other words, the non-zero $\RP$ values exhibited a diverse range.
As an example for illustration, among the five violations to MR1.1 for KM ($\VR = 5\%$ in Table~\ref{tb:VR}), we observed five diverse $\RP$ values (in ascending order): 0.55\%, 0.67\%, 0.93\%, 46.67\%, and 48.19\% (mean = 19.40\%). Table~\ref{tb:RP} shows the mean values of $\RP$ for the non-zero violation cases for each system with respect to each generic MR\@. Due to page limitation, the table shows the mean values of $\RP$ rather than their individual values.

\begin{table}[t]
	\renewcommand\arraystretch{1}
	\centering
	\caption{Mean Values of $\RP$ for Subject Clustering Systems with respect to Generic MRs} \label{tb:RP}
	\scalebox{0.8}{
		\begin{tabular}{cccccccc} 
			\toprule
			\zzy{Type of MRs} & MR & \multicolumn{3}{c}{Prototype-based} &\multicolumn{2}{c}{Hierarchy-based} & Density-based \\
			\cmidrule{3-5}  \cmidrule{6-8}
			& & KM & XM &EM &AN &FF & DS  \\
			\midrule
			\multirow{2}{0cm}{1} &
			1.1  & 19.40\% & 11.81\% & 0 & 0 & 7.40\% & 1.11\%  \\
			& 1.2  & 0 & 0 & N/A & N/A & 0 & N/A \\
			\hline
			\multirow{2}{0cm}{2} &
			2.1  & 49.77\% & 0 & 0 & N/A & 0 & N/A     \\
			& 2.2  & 23.36\% & 0 & 0 & 0 & 8.51\% & 16.31\%   \\
			\hline
			\multirow{2}{0cm}{3} &
			3.1 & 7.62\% & 35.75\% & 1.22\% & 1.92\% & 7.94\% & 2.74\% \\
			& 3.2 & 18.16\% & 17.28\% & 0.83\% & 1.67\% & 7.91\% & 4.48\% \\
			\hline
			\multirow{2}{0cm}{4} &
			4.1  & 46.34\% & 0 & 0 & 0 & 7.46\% & 0   \\
			& 4.2  &5.74\% & 0 & 0 & 0 & 0 & 0.64\%    \\
			\hline
			\multirow{2}{0cm}{5} &
			5.1 & 3.84\% & 2.37\% & 3.79\% & 16.15\% & 15.84\% & 8.48\% \\
			& 5.2 & 0 & 0 & 0 & 0 & 0 & 0 \\
			\hline
			\multirow{1}{0cm}{6} &
			6  & 11.53\% & 11.83\% & 1.99\% & 0 & 7.93\% & 1.97\% \\
			
			\bottomrule
	\end{tabular}}	
	\vspace{0.15cm}
	{\footnotesize
		\begin{flushleft} \dag~~%
			Each figure in the table denotes the mean value of $\RP$ over the violated trials with respect to the relevant MR.
		\end{flushleft}
	}
	\vspace{-0.1cm}		
\end{table}

Note that Tables~\ref{tb:VR} and~\ref{tb:RP} show the results in different perspectives. Table~\ref{tb:VR} counts the numbers of violated cases; while Table~\ref{tb:RP} focuses on the mean numbers of inconsistencies among those violated cases.
Also note that a high $\VR$ value does not necessarily imply a high $\RP$ value. Take FF under MR3.1 as an example. Here, reclustering occurred for 95 times among all the $100$ trials ($\VR = 95\%$). However, the mean percentage of reclustering was less than 8\% (mean number of $\RP$ $=$ 7.94\%). Thus, the results indicate that, although MR3.1 was often violated by FF, the extent of reclustering in these violations was quite marginal on average.
In contrast to FF, although XM violated MR3.1 only nine times ($\VR = 9\%$), this method had a mean value of $\RP$ of 35.75\%.  

\afterpage{
	\begin{landscape}
		\begin{figure}[htb]
			\centering
			\subfigure[KM]{\label{subfig:KMRPDistribution}
				\begin{minipage}[hbp]{0.65\textwidth}
					\includegraphics[width=1\textwidth, height=0.26\textheight]{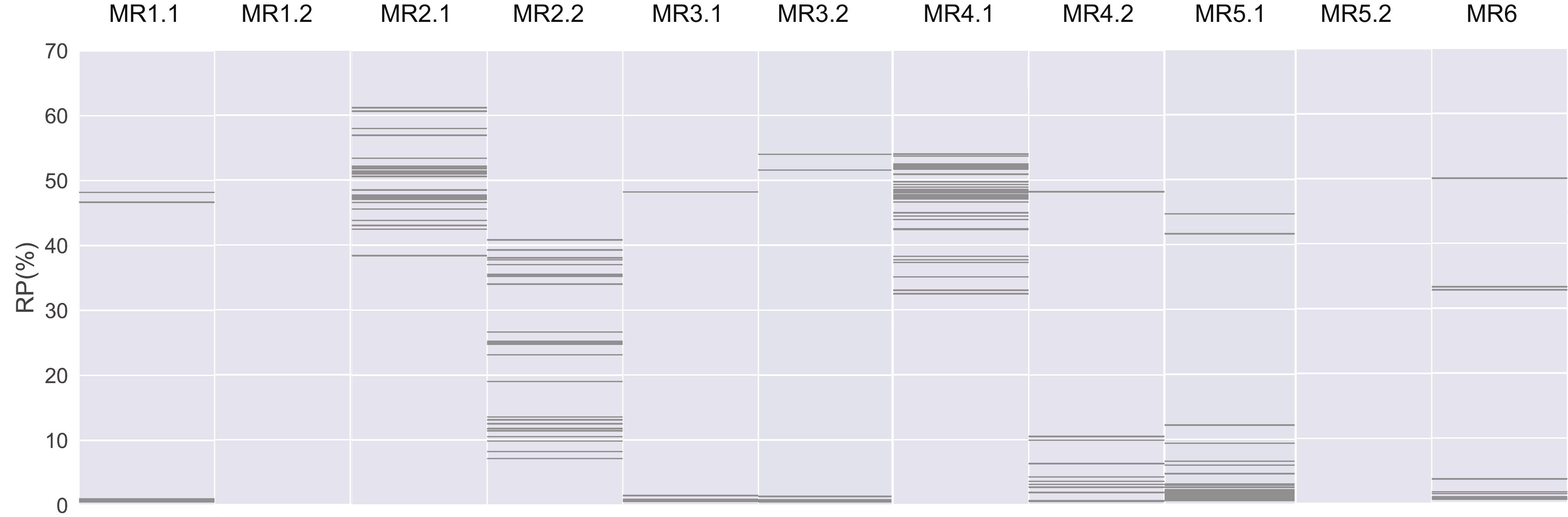}
			\end{minipage}}
			\subfigure[XM]{\label{subfig:XMRPDistribution}
				\begin{minipage}[hbp]{0.65\textwidth}
					\includegraphics[width=1\textwidth, height=0.26\textheight]{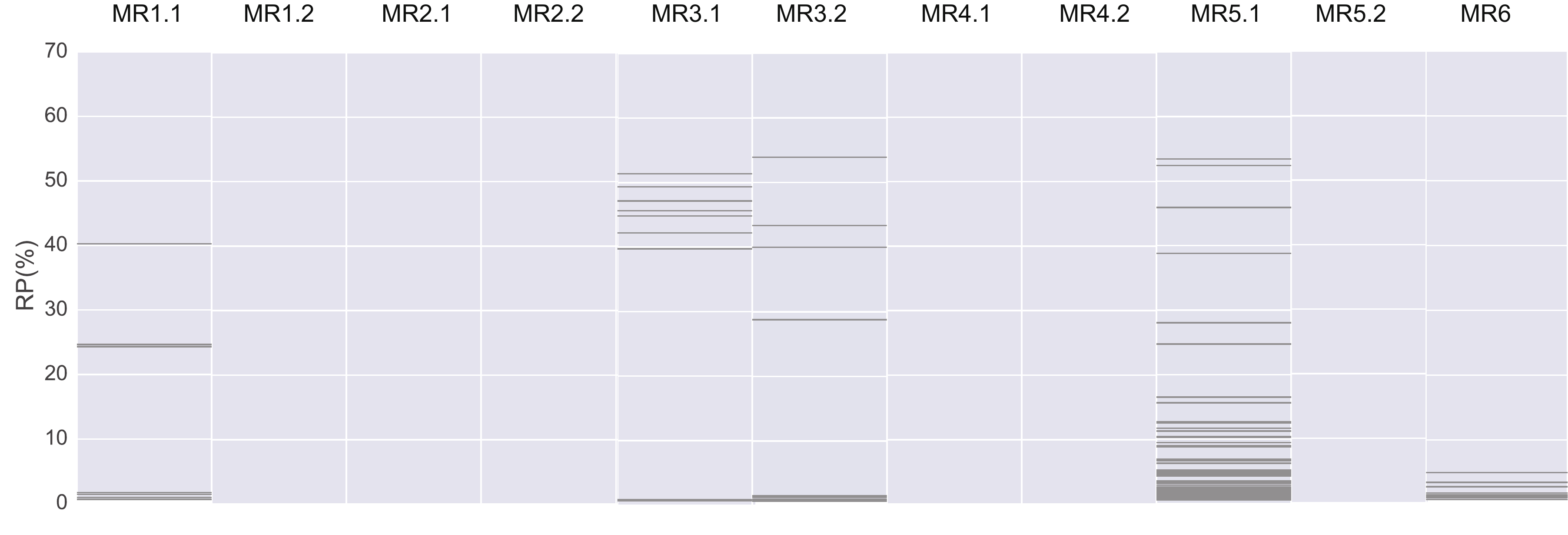} 
			\end{minipage}}
			\subfigure[EM]{\label{subfig:EMRPDistribution}
				\begin{minipage}[hbp]{0.65\textwidth}
					\includegraphics[width=1\textwidth, height=0.26\textheight]{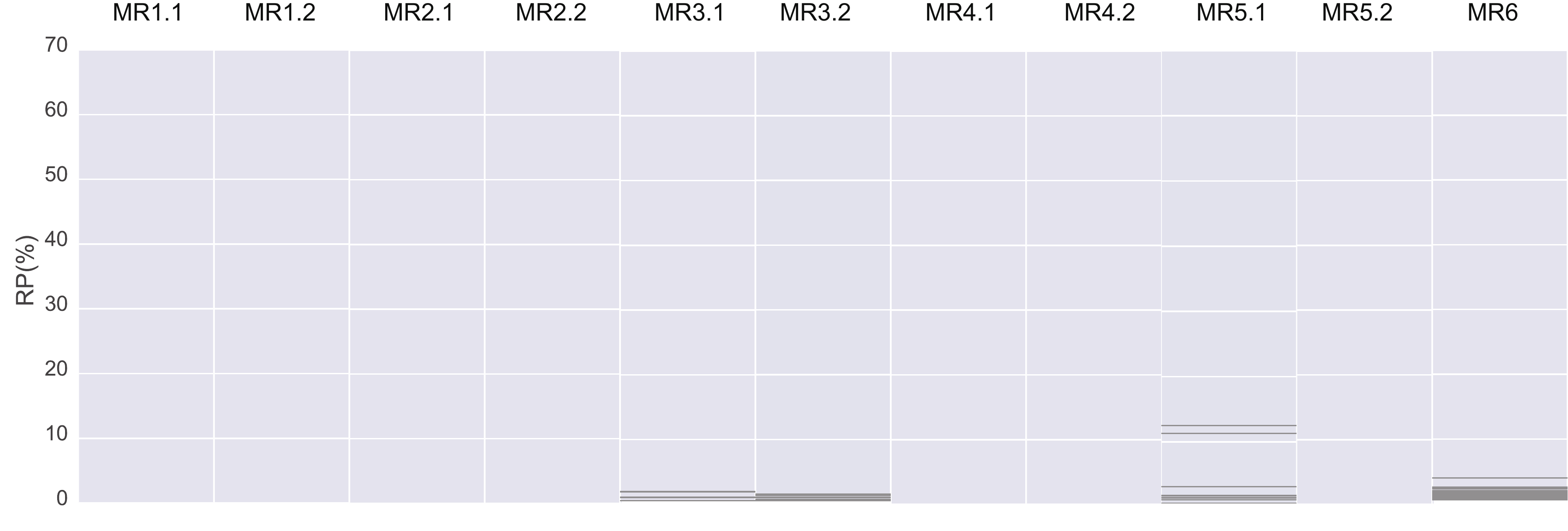} 
			\end{minipage}}
			\subfigure[AN]{\label{subfig:AGRPDistribution}
				\begin{minipage}[hbp]{0.65\textwidth}
					\includegraphics[width=1\textwidth, height=0.26\textheight]{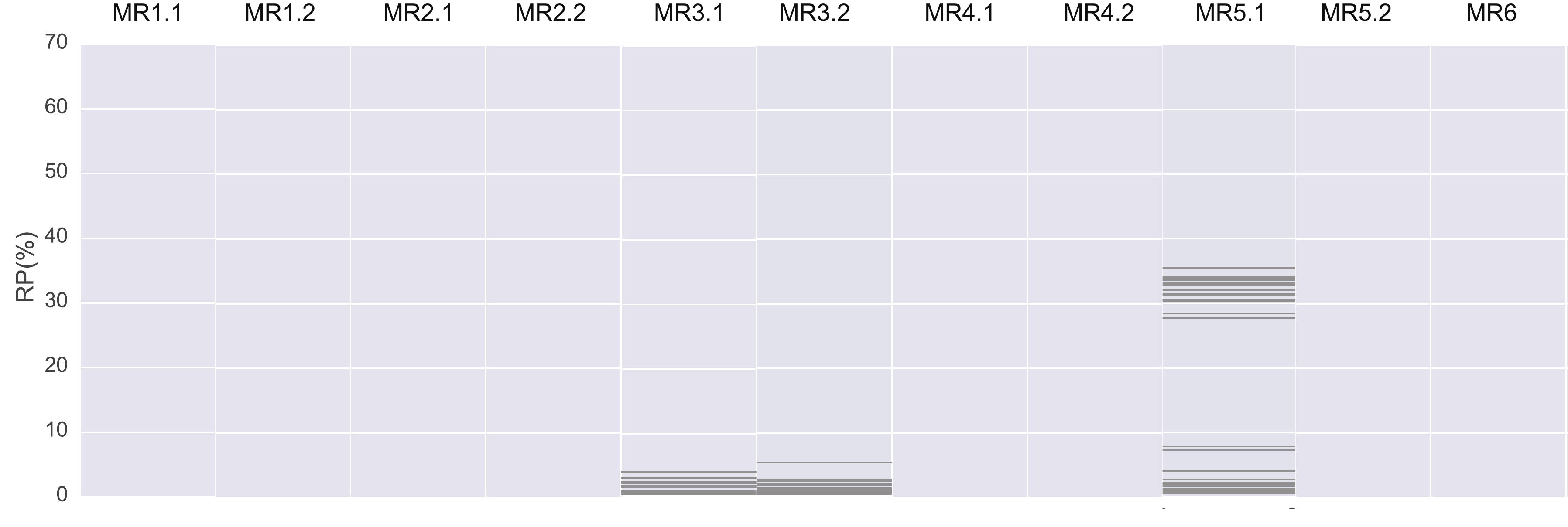} 
			\end{minipage}} 
			\subfigure[FF]{\label{subfig:FFRPDistribution}
				\begin{minipage}[hbp]{0.65\textwidth}
					\includegraphics[width=1\textwidth, height=0.26\textheight]{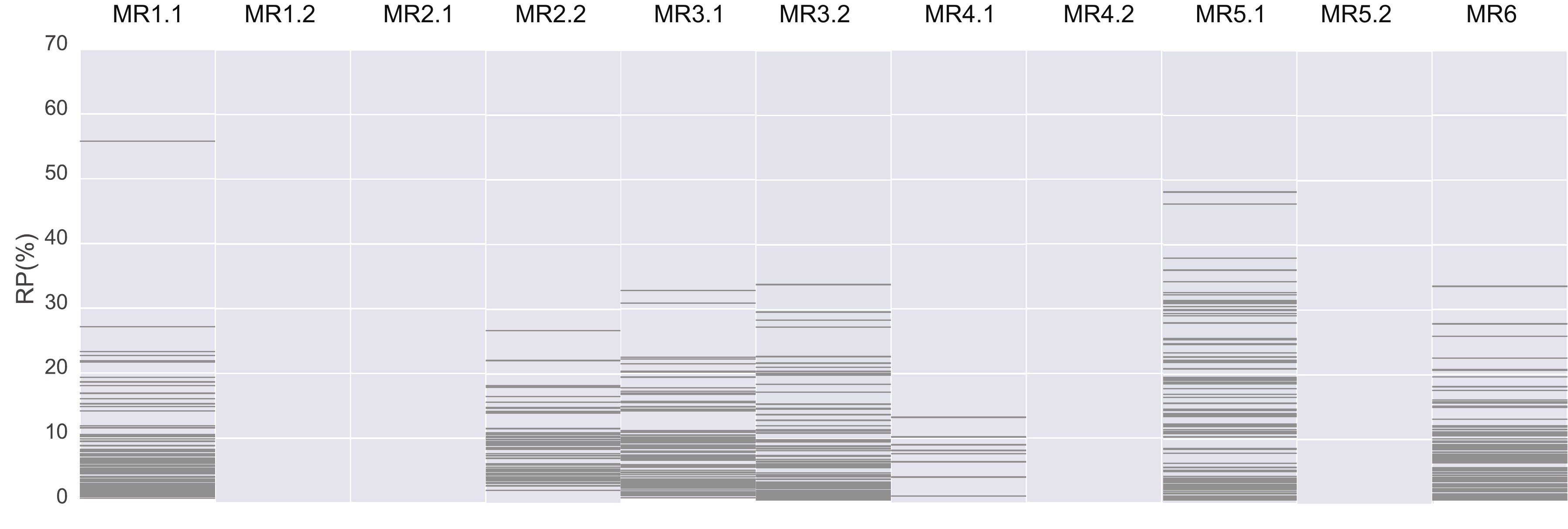} 
			\end{minipage}}
			\subfigure[DS]{\label{subfig:DBRPDistribution}
				\begin{minipage}[hbp]{0.65\textwidth}
					\includegraphics[width=1\textwidth, height=0.26\textheight]{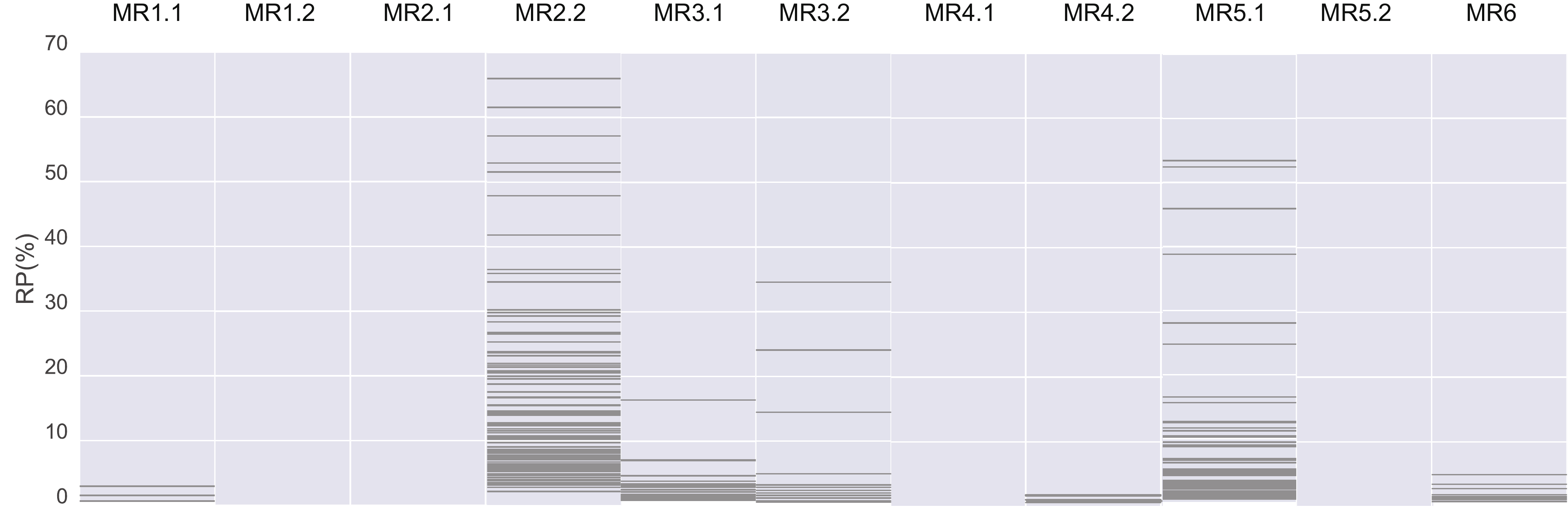} 
			\end{minipage}}
			\caption{$\RP$ distributions across the 11 generic MRs for each method.} \label{fig:RPDistribution}  
		\end{figure}
	\end{landscape}
}

We now turn to Fig.~\ref{fig:RPDistribution}, which combines the results in Tables~\ref{tb:VR} and~\ref{tb:RP} in one figure. In this figure, each horizontal bar corresponds to a violation to a particular MR by a system.
In each sub-figure, the largest value of $\RP$ shown in the y-axis is 70\%, because this was the largest $\RP$ value we observed across all the 11 MRs and all the six subject clustering systems in our experiment. In Fig.~\ref{fig:RPDistribution}, we can easily observe the "density" of the occurrences of reclustering over a certain range. Consider, for example, the set of horizontal bars related to MR2.2 and DS in Fig.~\ref{subfig:DBRPDistribution}. By looking at the distribution pattern of the horizontal bars, we know that the values of $\RP$ had a larger deviation in the higher-value ranges (closer to 70\%) than in the lower-value ranges (closer to zero percentage).  

Below we summarize the above findings:

\Sskip
\begin{itemize}
\item The 11 generic MRs have different capabilities to help a user detect "unexpected" behavior in clustering systems (from the user's perspective). More specifically:

\begin{itemize}
\Sskip
\item MR5.1 (related to the rotation of the coordinate system) is the most effective MR in identifying the corresponding "unexpected" behavior across all the six subject methods.

\Sskip
\item Some generic MRs, particularly MR5.2, could be necessary properties of clustering systems and, as such, no violation has been observed.
\end{itemize}

\Sskip
\item The robustness of handling each type of data transformation (as represented by the relevant generic MR) varied across the clustering systems, in terms of the $\VR$ and $\RP$ measures. More specifically:

\begin{itemize}
\item KM and FF had the worst performance across the 11 generic MRs.

\Sskip
\item On the other hand, EM and AN stayed relatively robust, yielding more desired results.
\end{itemize}

\end{itemize}

\vspace{-0.5ex}
\subsection{Underlying Behaviors that Cause Violations (RQ2)}
\label{sec:discussion}

This section complements Section~\ref{sec:overallresults} by drilling down to the \zzy{underlying behaviors and plausible reasons for the violations to each generic MR\@.}

\zzy{For each violation, we carefully inspected the results of both source and follow-up executions, by visualizing their clustering patterns. Such patterns are \ppl{fairly evident and immediate to users; these patterns} could easily and intuitively comprehend the anomalous behaviors of the subject clustering systems.}
Five types of clustering patterns were identified and shown in Table~\ref{tb:patterns}. (Note that Table~\ref{tb:patterns} only shows the pattern types that we observed in our experiment, rather than all the different possible pattern types.) In this table, each pattern type may be associated with more than one "similiar" pattern with non-identical data distributions and numbers of clusters. In what follows, we will illustrate the observed clustering pattern types and the underlying causes for their occurrence.

\begin{table*}[t]
	\renewcommand\arraystretch{1.3}
	\caption{Different Types of Clustering Patterns and their Related clustering systems}\label{tb:patterns}
	\centering
	\scalebox{1}{
		\begin{tabular}{lp{11cm}l} \toprule
			&  & Related \\
			Pattern type & Description & Clustering Systems \\
			\midrule
			BORDER  & Data objects near the boundary of one cluster in the source dataset are reassigned to \emph{different} clusters in the follow-up dataset. & All  \\
			MERGE \& SPLIT & Two source clusters are merged into one follow-up cluster, \emph{and} another source cluster is split into two smaller follow-up clusters. &  KM, FF \\
			SPLIT & One or more source clusters are split into smaller follow-up clusters. & EM, AN \\
			NOISE & Reclustering mainly occurs for those objects that are considered "noise''. & DS  \\
			NUM & The numbers of clusters differ between the results after the source and the follow-up executions. & DS \\
			\bottomrule
	\end{tabular}}
	\vspace{0.1cm}
\end{table*}

\Sskip
\subsubsection{Violations Related to KM and XM}
\label{subsec:km}

Figs.~\ref{subfig:KMRPDistribution} and~\ref{subfig:XMRPDistribution} show the distributions of the $\RP$ values for KM and XM, respectively, with respect to all the 11 generic MRs. For both systems, their $\RP$ values generally varied across a wide range (between $0\%-70\%$). For all the violations related to both systems, we took a close examination of the clustering results, and revealed two types of clustering patterns. In other words, these two pattern types occurred for both KM and XM\@. For the rest of Section~\ref{subsec:km}, to avoid lengthy discussion, we mainly discuss the results related to KM, followed by a short discussion on the results related to XM\@.

\begin{figure}[t]
	\centering
	\subfigure[Source dataset]{ \label{subfig:KMsource1}
		\includegraphics[width=0.22\textwidth]{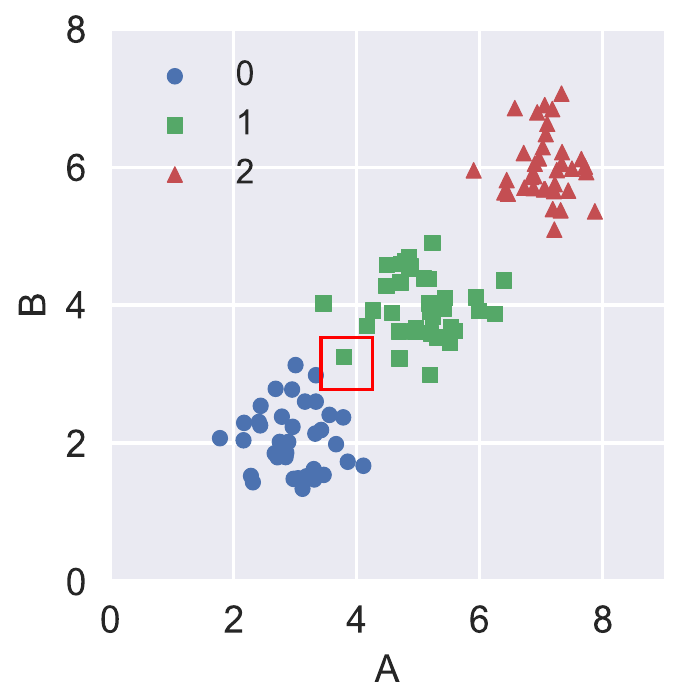}}
	\hspace{0.04in}
	\subfigure[Follow-up dataset]{ \label{subfig:KMfollow1}
		\includegraphics[width=0.22\textwidth]{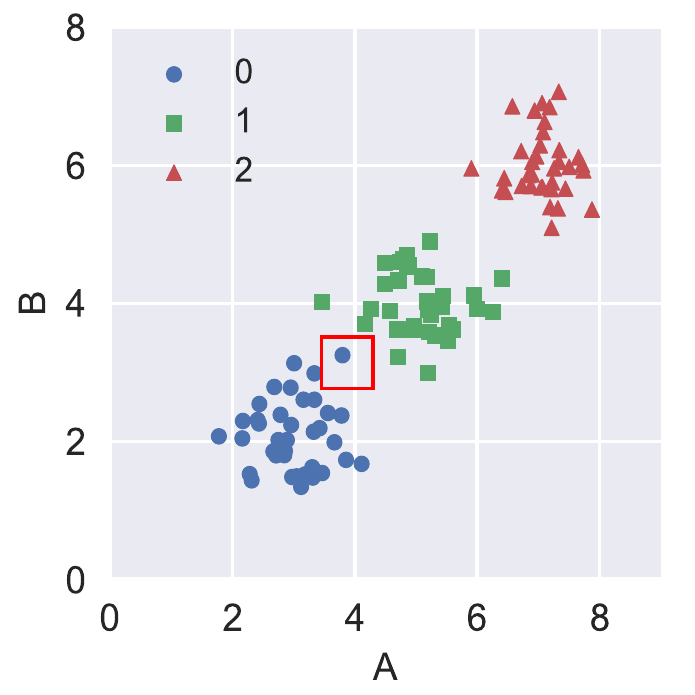}}
	\caption{Pattern type BORDER for KM\@. }\label{fig:KMSample1}
\end{figure}

\Nskip
\noindent
\textbf{BORDER\@.}\,
For those violations related to KM with relatively low $\RP$ values (e.g., $\RP<10\%$), some data points near the boundaries of clusters were reassigned to \emph{different} clusters in the follow-up dataset, as shown in Fig.~\ref{fig:KMSample1}. For simple illustration, this figure only shows one data point (enclosed in a \zzy{small} box) reassigned from one cluster (near its boundary) to an adjacent cluster. 
\zzy{(Note that the data point ``$\blacksquare$'' in the box in Fig.~\ref{fig:KMSample1}(a) has become the data point ``$\bullet$'' in the box in Fig.~\ref{fig:KMSample1}(b)).} However, in our experiment, more than one data points were observed to be reassigned to different clusters.

	\begin{figure}[t]
	\centering
	\subfigure[Source dataset]{ \label{subfig:KMsource2}
		\includegraphics[width=0.22\textwidth]{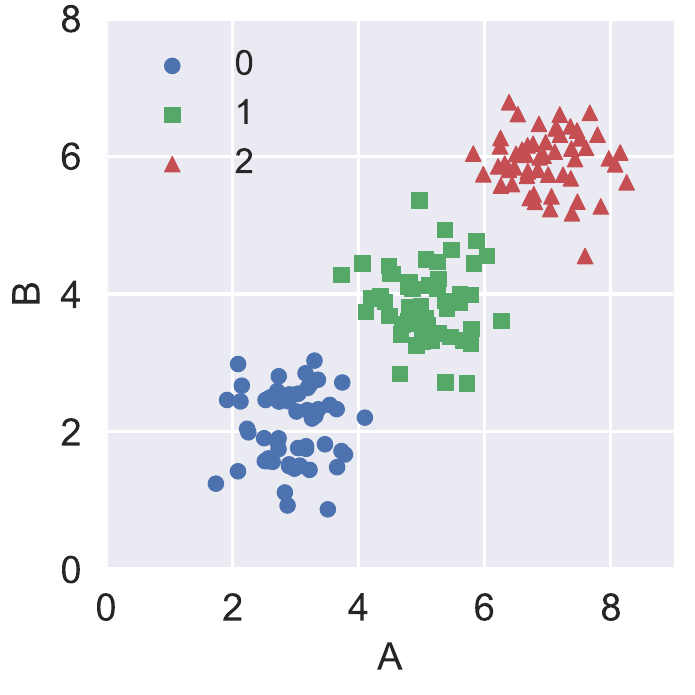}}
	\hspace{0.04in}
	\subfigure[Follow-up dataset]{ \label{subfig:KMfollow2}
		\includegraphics[width=0.22\textwidth]{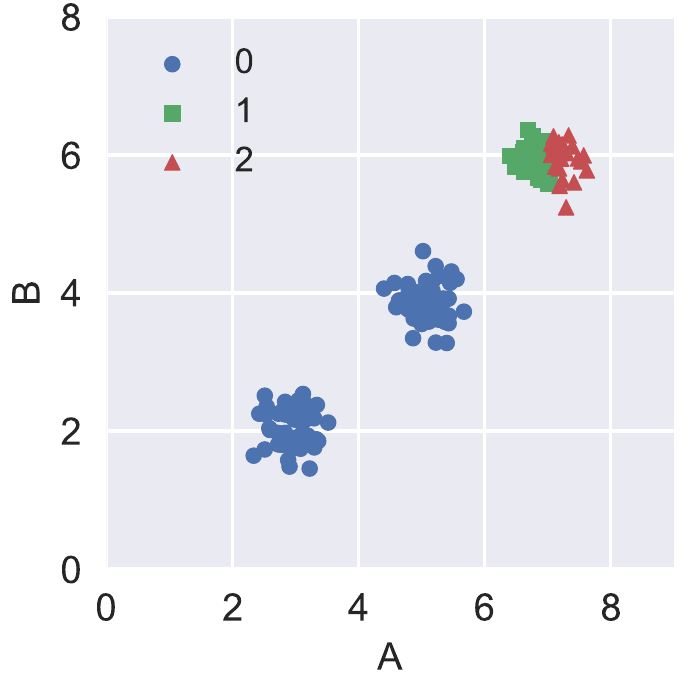}}
	\caption{Pattern type MERGE \& SPLIT for KM\@.}\label{fig:KMSample2}
\end{figure}

This pattern type was observed in the violations to MR1.1, MR3.1, MR3.2, MR5.1, and MR6. Some statistics on the violations related to BORDER and their $\RP$ values are provided as follows: 
(MR1.1) 60\% violations (3 out of 5, where $\RP{<}1\%$).
(MR3.1) 86\% violations (6 out of 7, where $\RP{<}1\%$).
(MR3.2) 67\% violations (4 out of 6, where $\RP{ < }1\%$).
(MR5.1) 96\% violations (55 out of 57, where $\RP{ < }10\%$).
(MR6) 73\% (8 out of 11, where $\RP{ < }5\%$). 
	
	Apparently, some users may think that KM is sensitive to the initialization condition (i.e., the selection of starting centroids). Thus, even a slight change on the starting centroids caused by data transformation (such as reordering or adding data samples) could lead to fairly different clustering results. Below we use MR1.1 (changing the object order) and MR5.1 (rotating the coordinate system) as examples to explain how data transformation affects the clustering results generated by KM\@.
	
	Consider MR1.1 first. Reordering data samples has no effect on data distribution, but is likely to change the randomly initialized (starting) cluster centroids. We argue that, with the gradual relocation of cluster centroids following each iteration of reclustering, KM may finally generate a different set of data clusters. Our argument was validated by MR1.2 that no violation occurred if we changed the object order but keeping the same set of starting centroids.
	With respect to MR1.1, we carefully checked the clustering process and confirmed that the starting centroids in the \emph{violated} trials were actually changed after changing the object order. But, at the same time, we also observed that many \emph{non-violated} trials involved changing their starting centroids.
	Therefore, the results have suggested that KM may not be as sensitive to the starting centroids as some users initially conceive.
	
	Next, we turn to MR5.1. Many users of KM generally expect that rotating the coordinate system will not affect the clustering result, because such rotation does not change the data distribution pattern.
However, this was not the case observed in our experiment; we found some "unexpected" violations to MR5.1.
By inspecting the source code of KM collected from Weka, we found a function $\mathit{distance}$ for calculating the Euclidean distance between an arbitrary object $\textbf{x}_i$ and each cluster centroid. 
Before executing the core part of the distance computation, KM normalizes each attribute value with min-max normalization via the function $\mathit{norm}$. As such, the centroid $\textbf{m}_k$ nearest to $\textbf{x}_i$ will be chosen and $\textbf{x}_i$ will be assigned with label $k$. By checking the output after each iteration of KM, we found that the normalized Euclidean distance between $\textbf{x}_i$ and $\textbf{m}_k$ was different between the source and the follow-up executions, although the theoretical distance remains unchanged after rotating the coordinates. Hence, a small change on the distance could result in a different decision by KM when choosing the nearest centroid. Furthermore, the impact of min-max normalization will be brought forward into subsequent iterations, thereby explaining the major reason for violating MR5.1.

\Nskip	
\noindent
\textbf{MERGE \& SPLIT\@.}\,
	Most KM-related violations with their $\RP$ values larger than 10\% were associated with this pattern type (see Fig.~\ref{fig:KMSample2} for an example). For the MERGE \& SPLIT pattern type, two source clusters are merged into one follow-up cluster, and one other source cluster is split into two smaller follow-up clusters.

This pattern type was associated with all the violations, which were related to all the 11 generic MRs except MR1.2 and MR5.2.
Some statistics on the violations related to MERGE \& SPLIT and their $\RP$ values are provided as follows: 
(MR1.1) 40\% violations (2 out of 5, where $\RP{ > }40\%$);
(MR2.1) 100\% violations (26 out of 26, where $\RP{ > }38\%$);
(MR2.2) 100\% violations (35 out of 35, where $\RP{ > }10\%$);
(MR3.1) 14\% violations (1 out of 7, where $\RP{ > }40\%$);
(MR3.2) 33\% violations (2 out of 6, where $\RP{>}40\%$);
(MR4.1) 100\% violations (36 out of 36, where $\RP{>}30\%$);
(MR4.2) 6\% violations (1 out of 17, where $\RP{>}40\%$);
(MR5.1) 4\% violations (2 out of 57, where $\RP{>}40\%$);
(MR6) 27\% violations (3 out of 11, where $\RP{>}30\%$).

	It is commonly known that KM may quickly converge to a local optimum, resulting in unsatisfactory results. We conjecture that the MERGE \& SPLIT pattern type occurred due to this reason. 
	To test this conjecture, we compared the iteration numbers between the source and follow-up executions. Our rationale is based on the intuition that a low iteration number (i.e., an early iteration) is normally associated with high convergence speed, and high convergence speed is often a signal of prematurity, resulting in a local optimum.

\begin{figure}[t]
		\centering
		\subfigure[Histogram of $\SFR$ values related to MR2.1 (100 trials).]{\label{subfig:histogram}
			\includegraphics[width=0.4\textwidth]{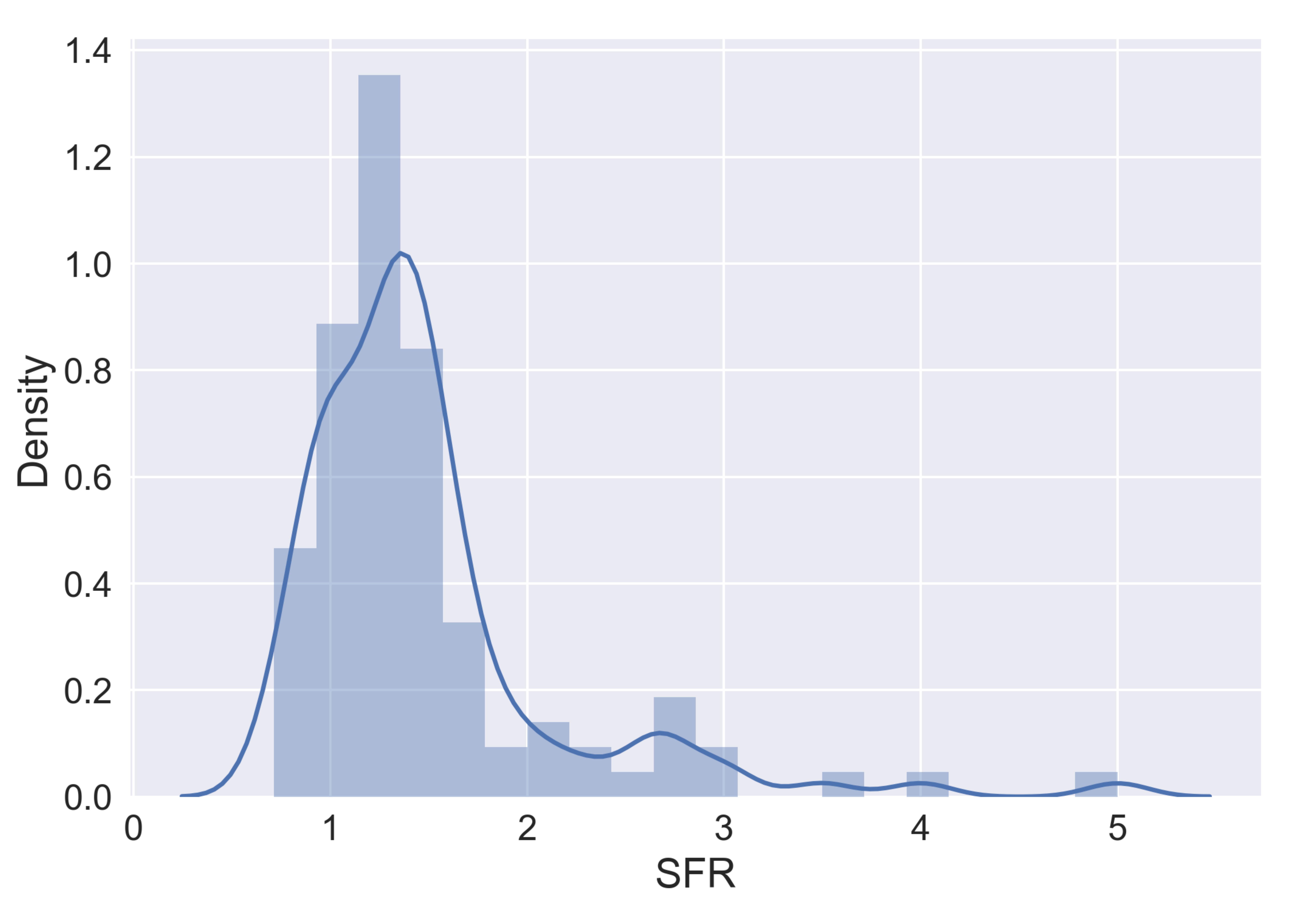}}
		\subfigure[Distributions of $\RP$ and $\SFR$ values related to MR2.1 (100 trials).]{\label{subfig:bubblegram}
			\includegraphics[width=0.45\textwidth]{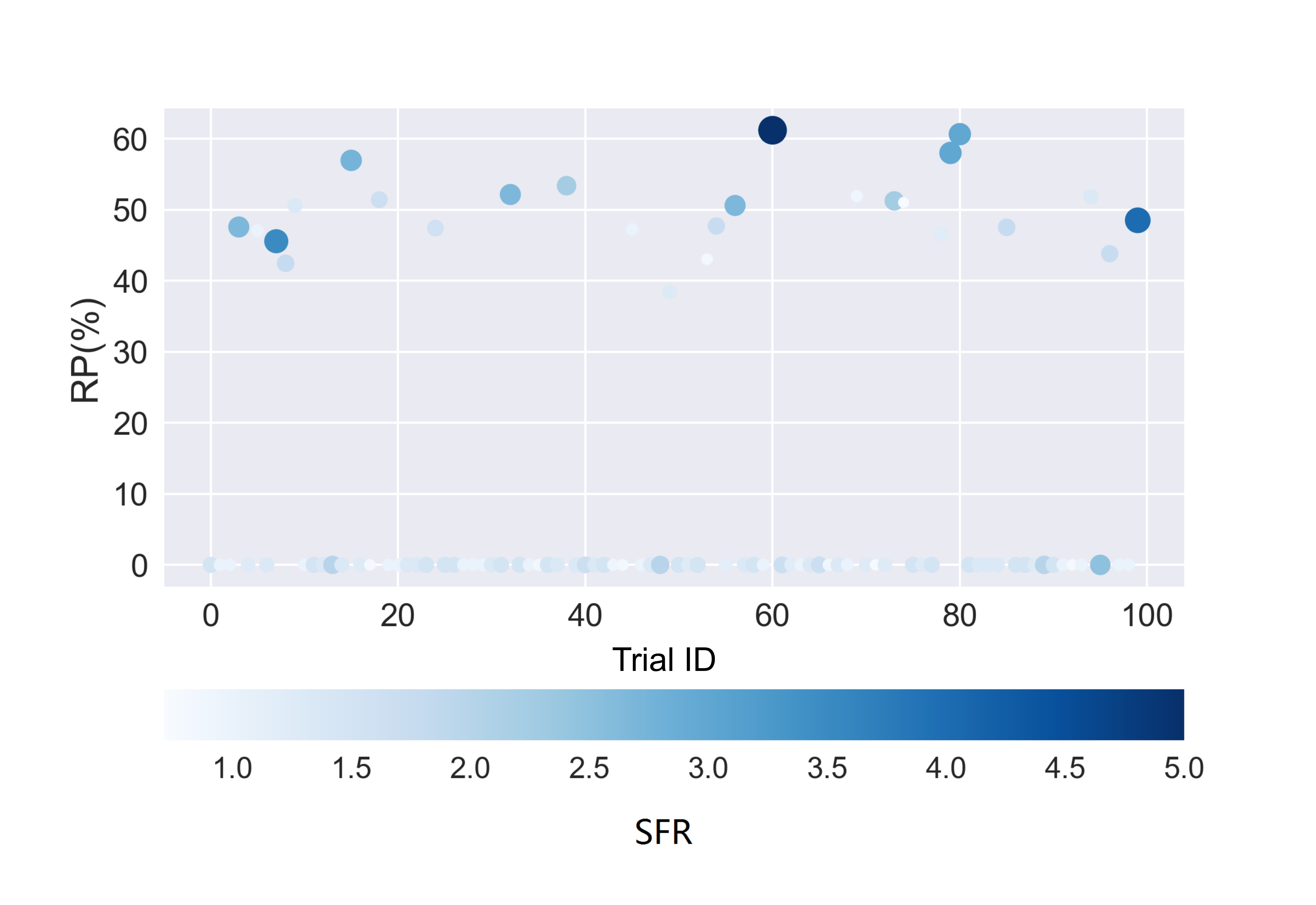}}
		\caption{Distributions of $\SFR$ and $\RP$ values related to MR2.1 (100 trials).} \label{fig:iteration}
	\end{figure}

	Here, we use MR2.1 as an example for illustration: If a set of data samples can be well-clustered, then shrinking each cluster towards its centroid should make the clusters more compact, thereby producing an even more clearcut clustering result.
	Let $I_s$ and $I_f$ denote the iteration numbers in the source and follow-up clustering processes, respectively. Let ${\SFR}\,=\,\frac{I_s}{I_f}$. Obviously, ${\SFR}\,{>}\,1$ indicates less iterations and a higher convergence speed in the follow-up clustering process; while ${\SFR}\,{ < }\,1$ indicates the opposite situation.
	Fig.~\ref{subfig:histogram} illustrates the distribution of $\SFR$ values related to MR2.1 for $100$ trials in a histogram. 	
	From this figure, we observed that among the 100 trials, $10\%$ of them had their $\SFR$ values less 1.0, and $19\%$ of them had their $\SFR$ values equal to 1.0. Among the remaining 71\% of the trials whose $\SFR > 1$, 79\% have $1 < {\SFR} \leq 2$, 18\% have $ 2 < {\SFR} \leq 3$, and 3\% have ${\SFR} > 3$.
	
	The main upper portion of Fig.~\ref{subfig:bubblegram} shows the distribution of $\RP$ values related to MR2.1 over $100$ trials. Each dot at position $(i,\,j)$ in the figure indicates that the $i$th trial had its corresponding ${\RP}\,{=}\,j$. 
	Note that the size and the darkness of the round dots are proportional to their $\SFR$ values: the larger and darker a dot is, the higher is its corresponding $\SFR$ value (and, hence, the higher is the convergence speed in the follow-up clustering process).

The horizontal bar at the bottom of Fig.~\ref{subfig:bubblegram} indicates the value ranges of $\SFR$.
According to the definition of $\RP$, ${\RP}\,{=}\,0\%$ indicates no violation to the relevant MR, while ${\RP}\,{>}\,0\%$ indicates the existence of a violation. In all the violated cases related to MR2.1, data were clustered in patterns similar to Fig.~\ref{fig:KMSample2}, resulting in fairly high $\RP$ values. 
It can be seen from Fig.~\ref{subfig:bubblegram} that almost all trials with ${\RP}\,{=}\,0\%$ had their $\SFR$ values close to 1 (see the small and light dots on the horizontal line (i.e., those that are parallel and just above the x-axis) corresponding to ${\RP}\,=\,0\%$), indicating that the source and the follow-up processes had similar convergence speeds. 
On the other hand, those trials with very high $\RP$ values were most likely associated with high $\SFR$ values (see the large and dark dots on the horizontal line (i.e., those that are parallel and just above the x-axis in Figure~\ref{subfig:bubblegram}), indicating that their follow-up processes were faster than the source processes. In particular, the large and dark dot for the trial ID 60 corresponds to a violated trial with ${\RP}\,>\,60\%$, and its follow-up process was about four times faster than its corresponding source process.
Figure~\ref{subfig:bubblegram} also shows a positive correlation between $\RP$ and $\SFR$ values with respect to MR2.1\@.
All the above analyses have demonstrated that, with respect to KM, violations with high reclustering percentages were very likely due to an accelerated convergence to local optima.
For other MRs with MERGE \& SPLIT\@ violation pattern in KM, we observed similar phenomena.

We now turn to XM\@. In terms of the violations to MR1.1, MR3.1, MR3.2, MR5.1, and MR6, XM was not better than KM\@.
For these five MRs, the clustering pattern types observed for KM also occurred for XM\@. Thus, we do not repeat the discussion on the violations related to XM\@. However, we would like to point out that,
when comparing with KM, XM was relatively more robust to the type of data transformation related to MR2.1, MR2.2, MR4.1, and MR4.2. A close examination on those violations related to these four MRs revealed that a common property existed, that the resulting clusters were relatively more clear-cut (for MR2.1) or separated from each others (for MR4.1). 

As an extension to KM, XM proposes a partial remedy for the local optimum problem \citep{pelleg2000x}. Many people argue that XM is less sensitive to local optima by searching for the true number of clusters in a predefined range. This argument was validated to be valid by {\sc mettle}\,---\,XM outperformed KM in the situations where data groups were largely separated. On the other hand, in those situations where data groups were well clustered but with a lower degree of separation, XM and KM generated similar clustering results. 
	
	\Sskip
	\begin{tcolorbox}[colframe=black!75!white, colback=white, boxrule=0.3mm]
		\textbf{\emph{Summary:}}\, The sensitivity of KM and XM to initial conditions and noisy data was validated by our experiment. Data transformation, such as reordering data and adding noises, will result in reassigning data objects near the boundary of one cluster to another cluster, which is normally expected by users. Our experiment also revealed an important property of KM: this system tends to converge to local optima even when the clusters are sufficiently well separated, which leads to high reclustering percentages. Although XM is theoretically less sensitive to local optima than KM, our experiment results show that XM only outperforms KM when the original dataset is highly separated.
	\end{tcolorbox}
	
\Sskip
\subsubsection{Violations Related to EM}
\label{subsec:EM}

\begin{figure}[t]
	\centering
	\subfigure[Source dataset]{ \label{subFig:MR3.1EMsource}
		\includegraphics[width=0.22\textwidth]{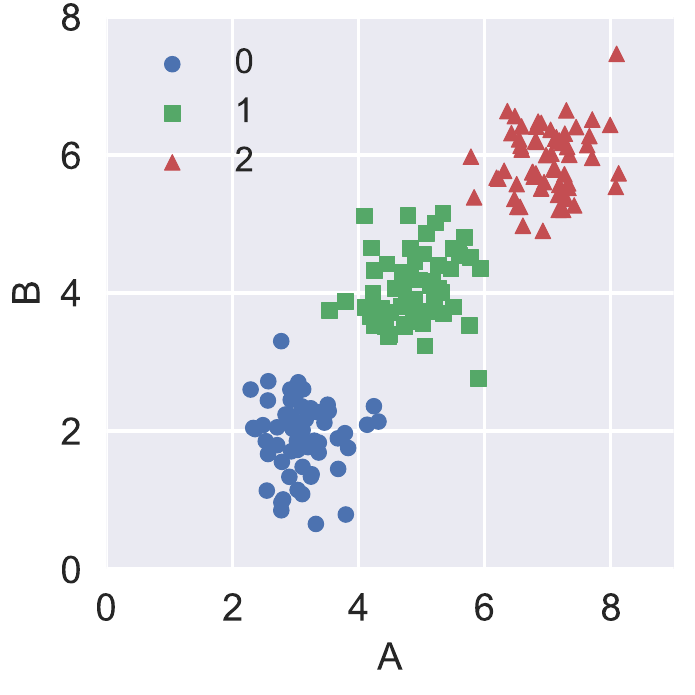}}
	\subfigure[Follow-up dataset]{ \label{subFig:MR3.1EMfollow}
		\includegraphics[width=0.22\textwidth]{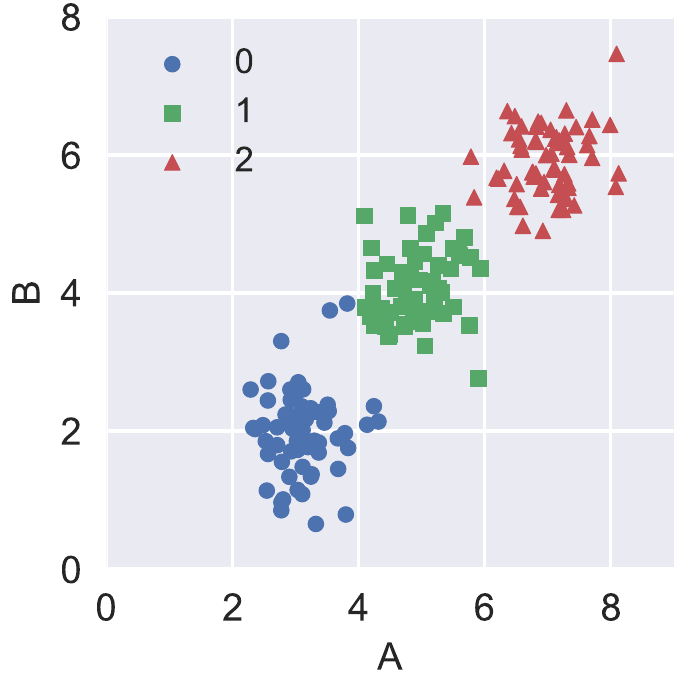}}
	\caption{Pattern type BORDER for EM\@.}
	\label{fig:MR3.1EMSample}
\end{figure}

\begin{figure}[t]
	\centering
	\subfigure[Source dateset]{ \label{subFig:MR5.1EMsource}
		\includegraphics[width=0.22\textwidth]{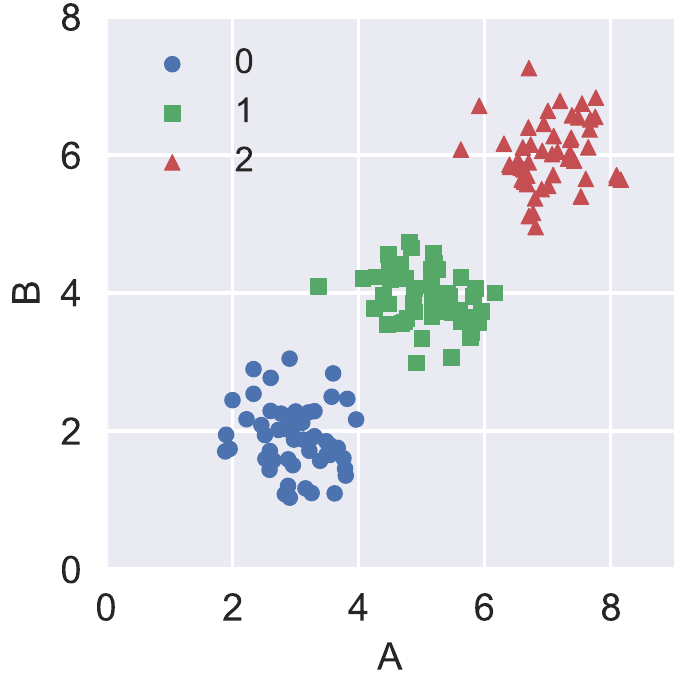}}
	\subfigure[Follow-up dataset]{ \label{subFig:MR5.1EMfollow}
		\includegraphics[width=0.22\textwidth]{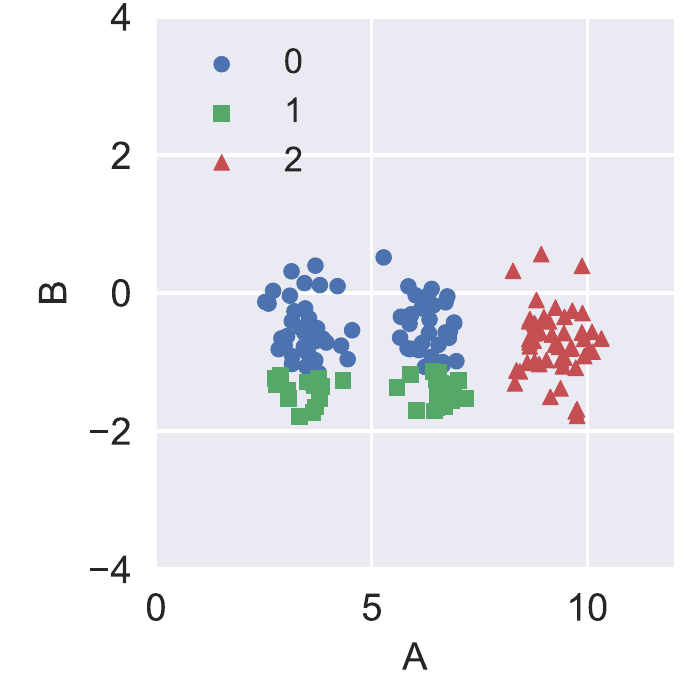}}
	\caption{Pattern type SPLIT for EM\@.}
	\label{Fig:MR5.1EMSample}
\end{figure}

Fig.~\ref{subfig:EMRPDistribution} shows that, for EM, violations only occurred in those cases related to MR3.1, MR3.2, MR5.1, and MR6. Among the 100 trials, the numbers of violated cases were 5, 11, 9, and 39, respectively, for these four MRs. Each of these violations had a low $\RP$ value, indicating that very few data samples were reassigned from one cluster to another. Based on these results, we argue that EM is fairly robust to different types of data transformation. We also found two clustering pattern types: BORDER and SPLIT\@.

\Nskip
\noindent
\textbf{BORDER\@.}\, We observed from Fig.~\ref{subfig:EMRPDistribution} that most of the violations related to EM had fairly low $\RP$ values. As shown in Fig.~\ref{fig:MR3.1EMSample}, only several data samples near the boundaries of clusters were reassigned to other clusters by EM, and this clustering result was consistent with many users' expectation.

This pattern type was observed in those violations to MR3.1, MR3.2, MR5.1, and MR6. Some statistics on the violations related to this pattern type and their $\RP$ values are provided as follows:
(MR3.1) 100\% violations (5 out of 5, where $\RP < 2\%$);
(MR3.2) 100\% violations (11 out of 11, where $\RP < 1\%$);
(MR5.1) 78\% violations (7 out of 9, where $\RP < 3\%$);
(MR6) 100\% violations (39 out of 39, where $\RP < 3\%$).

The above statistics indicate that, although the clustering results generated by EM was affected by the types of data transformation corresponding to MR3.1, MR3.2, MR5.1, and MR6, the impact on the clustering results was fairly small (as shown by the very small $\RP$ values). Also, Figure~\ref{fig:RPDistribution} shows that EM has the second smallest number of violated MRs ($= 4$) among all the six subject clustering systems. In this regard, EM has the best performance among the subject clustering systems according to the user's expectations (which are expressed in terms of the 11 generic MRs).  
	
One issue is worth mentioning here. Similar to KM and XM, violations to MR5.1 (rotating the coordinate system) were also observed for EM\@. As we have pointed out in Section~\ref{subsec:km}, violations to MR5.1 (and also other generic MRs) have revealed a gap between the actual performance and the user's expectation about a method (in this case, EM).
In the Weka implementation, EM initializes estimators by running KM $10$ times and then choosing the "best" solution with the smallest squared error for all the clusters. This chosen solution then serves as the basis for executing the E-step and the M-Step in EM (see Section~\ref{subsec:prototype}). Due to this reason, the clustering result generated by EM partially depends on KM, therefore it is not surprising to see that both EM and KM showed violations to MR5.1. 
	
\Nskip
\noindent
\textbf{SPLIT\@.}\, Fig.~\ref{Fig:MR5.1EMSample} shows an example of this pattern type: in the source dataset, each of the \zzy{two clusters (the cluster with the "$\bullet$" data points and the cluster with the "$\blacksquare$" data points)} clusters was split into two smaller clusters in the follow-up dataset; at the same time, merging of clusters in the source dataset did not occur.

This pattern type was only discovered in two out of the nine violations ($=$\,22\%) to MR5.1, with their $\RP$ values over $10\%$. As explained above, EM partially depends on the KM solution. Thus, violations to MR5.1 by EM could occur after rotating the coordinates. After a close examination, we found that the theoretically "best'' KM solution was not always what users normally expected. With respect to SPLIT, the chosen KM solution at the initialization stage was found to involve unexpected data partitions which was similar to the pattern type shown in Fig.~\ref{subFig:MR5.1EMfollow}. This explains why EM generated poor clustering results after iterations based on the ill-initialization by KM\@.

\Nskip	
	\begin{tcolorbox}[colframe=black!75!white, colback=white, boxrule=0.3mm]
		\emph{\textbf{Summary:}}\, According to the 11 generic MRs, EM is the most robust one among the six subject clustering systems. Reassigning data samples from one cluster to another cluster still occurred, which contradicted the user's expectation.
       However, since the $\RP$ values were very small ($< 3\%$), the impact of data transformation on the clustering result was much less than the other five subject clustering systems.
       Although both EM and KM execute in an iterative manner, our experiment shows that EM is less sensitive to local optima than KM\@. Furthermore, in Weka implementation, the theoretically "best'' solution chosen by EM during initialization may not be in line with the user's expectation, resulting in the poor clustering result generated by EM\@.
	\end{tcolorbox}
	
\Sskip
\subsubsection{Violations Related to AN}
\label{subsec:agnes}

It can be seen from Fig.~\ref{subfig:AGRPDistribution} that AN only caused violations to three generic MRs: MR3.1, MR3.2, and MR5.1. The $\RP$ values associated with MR3.1 and MR3.2 were very low ($< 5\%$ for both MRs). On the other hand, among all the $\RP$ values associated with MR5.1, some were under 10\% but the others were fairly high ($> 30\%$).
We found two clustering pattern types from those violations related to AN\@.

\begin{figure}[t]
	\centering
	\subfigure[Source dataset]{ \label{subFig:MR3.1AGsource}
		\includegraphics[width=0.22\textwidth]{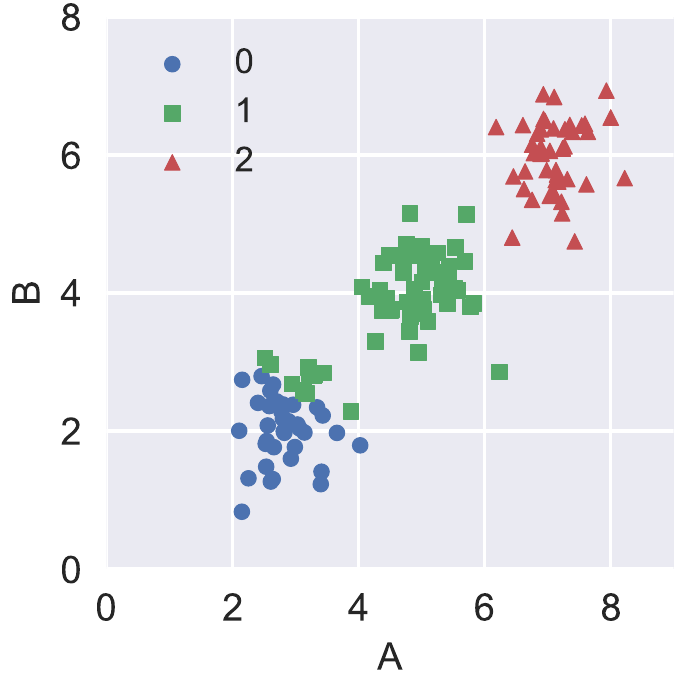}}
	\subfigure[Follow-up dataset]{ \label{subFig:MR1.1AGfollow}
		\includegraphics[width=0.22\textwidth]{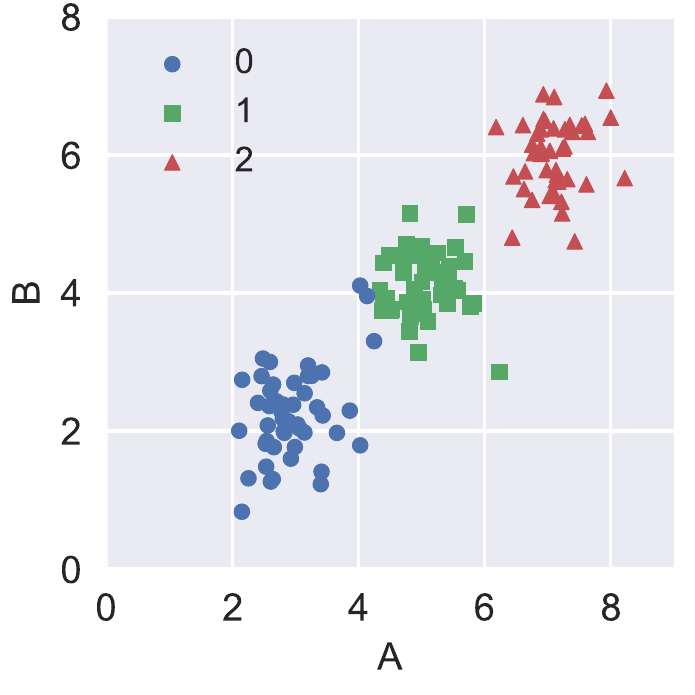}}
	\caption{Pattern type BORDER for AN.}
	\label{fig:MR3.1AGSample}
\end{figure}

\Nskip
\noindent
\textbf{BORDER\@.}\, For those violations with low $RP$ values ($< 10\%$), the same clustering pattern type as shown in Fig.~\ref{fig:MR3.1AGSample} was observed. For these violations, only several data samples near the cluster boundaries were affected. BORDER was observed in all violations to MR3.1 and MR3.2, and in 43\% (39 out of 91) violations to MR5.1\@.

\begin{figure}[t]
		\centering
		\subfigure[Source dataset]{ \label{subFig:MR5.1AGsource}
			\includegraphics[width=0.22\textwidth]{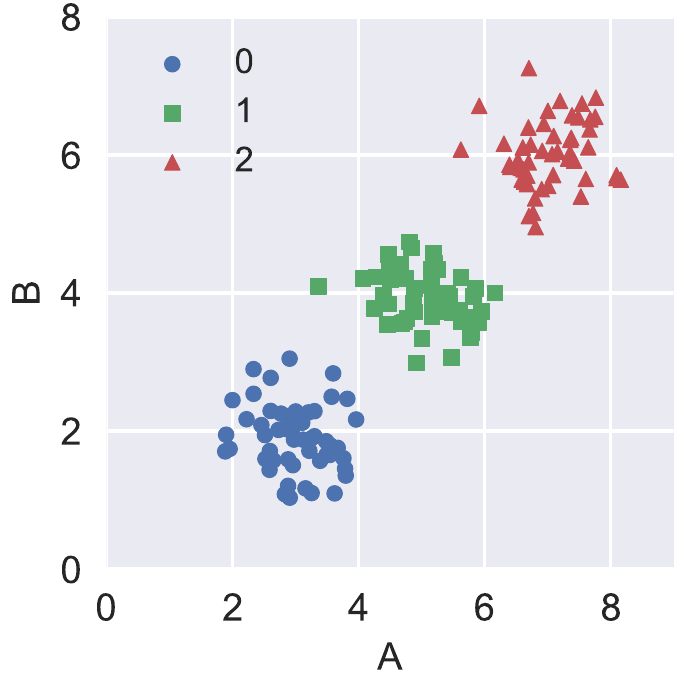}}
		\subfigure[Follow-up dataset]{ \label{subFig:MR5.1AGfollow}
			\includegraphics[width=0.22\textwidth]{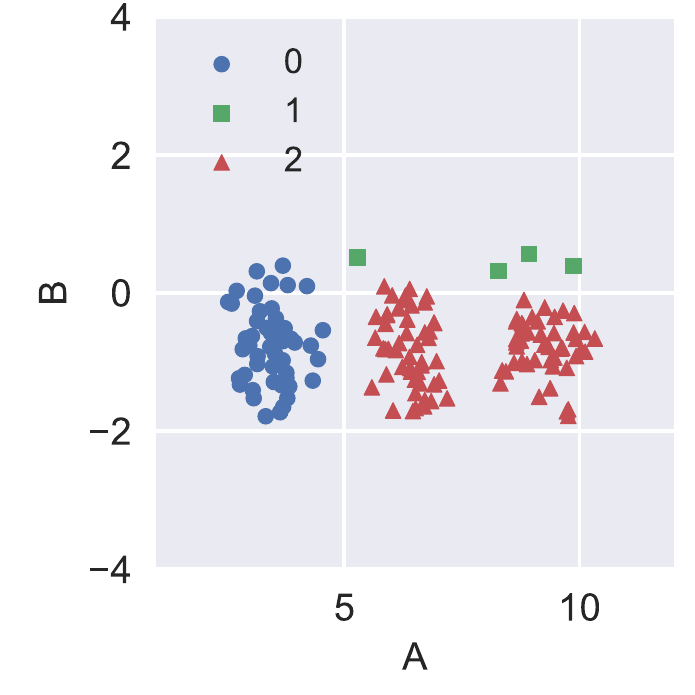}}
		\caption{Pattern type SPLIT for AN.}
		\label{Fig:MR5.1AGSample}
	\end{figure}

\Nskip
\noindent
\textbf{SPLIT\@.}\, For those violations with relatively high $\RP$ values ($> 30\%$),
we observed this pattern type (similar to the one shown in Fig.~\ref{Fig:MR5.1AGSample}), where each of the 
\zzy{two clusters (the cluster with the "$\blacksquare$" data points and the cluster with the "$\blacktriangle$" data points)}
in the source dataset was split into a small cluster (\zzy{with the ``$\blacksquare$" data points}) and a much larger cluster (\zzy{with the ``$\blacktriangle$'' data points}) in the follow-up dataset. 
After checking the Weka implementation, we found that min-max normalization is also adopted in the preprocessing phase of AN, causing the violations to MR5.1\@.
	
\Nskip
\begin{tcolorbox}[colframe=black!75!white, colback=white, boxrule=0.3mm]
	\emph{\textbf{Summary.}}\,  As a hierarchy-based clustering system, AN is more robust to data transformation when compared with FF\,---\,only boundary points are occasionally affected. Our experiment also revealed that, similar to other systems, there exists a gap between the performance of AN and the user's expectation on this system.
\end{tcolorbox}

\Sskip
\subsubsection{Violations Related to FF}
\label{subsec:FF}

Fig.~\ref{subfig:FFRPDistribution} shows that FF caused relatively more violations to the generic MRs when compared with other clustering systems, with the $\RP$ values ranged from 0\% to 50\%. We observed two clustering pattern types for FF.

\begin{figure}[t]
	\centering
	\subfigure[Source dataset]{ \label{subfig:MR1FFsource}
		\includegraphics[width=0.22\textwidth]{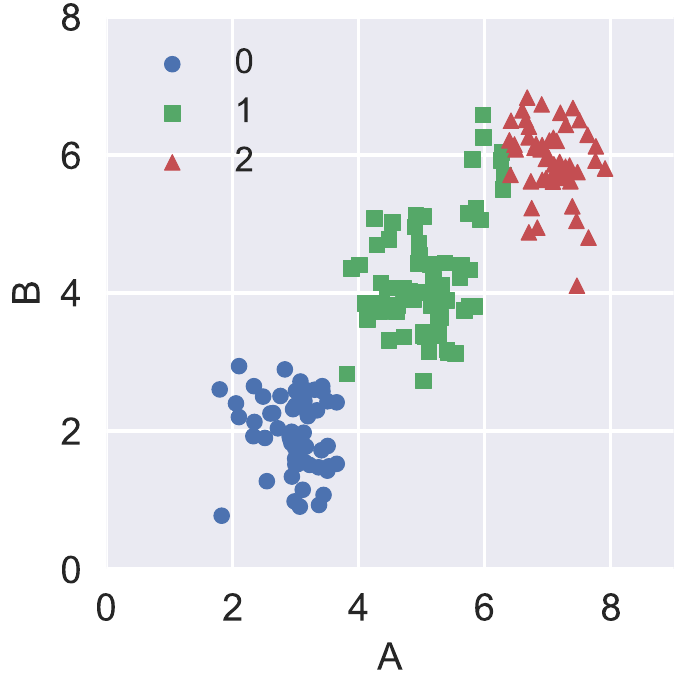}}
	\hspace{-0.12in}
	\subfigure[Follow-up dataset]{ \label{subfig:MR1FFfollow}
		\includegraphics[width=0.22\textwidth]{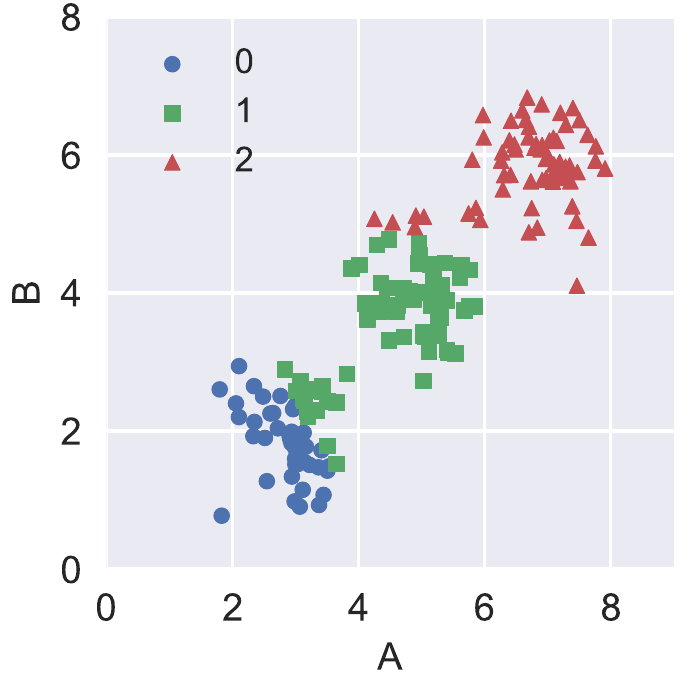}}
	\caption{Pattern type BORDER for FF\@.}\label{fig:MR1FFSample}
\end{figure}

\Nskip
\noindent
\textbf{BORDER\@.}\,
	For the violations where $\RP < 30\%$, data samples near the cluster boundaries were reassigned to different clusters, as shown in Fig.~\ref{fig:MR1FFSample}. This pattern type appeared for MR1.1, MR2.2, MR3.1, MR3.2, MR4.1, MR5.1, and MR6\@. 
When compared with other methods, the reclustering of data samples with this pattern type was not very precise. Consider, for example, in Fig.~\ref{fig:MR1FFSample}, a few data samples near the boundary of the cluster 
\zzy{with the "$\blacksquare$" data points}
in the source dataset were incorrectly reassigned to the 
cluster 
\zzy{with the "$\blacktriangle$" data points} in the follow-up dataset.
	
It can be seen from Table~\ref{tb:VR} that FF caused many violations to MR1.1, with $\RP\,=\,90\%$. This result supports our analysis on FF, that the clustering process and result are largely affected by the starting centroids chosen by FF~\cite{dasgupta2005performance}.
If the starting centroids selected by FF are changed by reordering the object order (MR1.1), the farthest-first traversal sequence may be affected. In addition, the fact that no violation to MR1.2 (this MR involves keeping the same set of starting centroids unchanged) was detected for FF further supports our analysis.

	\begin{figure}[t]
	\centering
	\subfigure[Source dataset]{ \label{subFig:MR5.1FFsource}
		\includegraphics[width=0.22\textwidth]{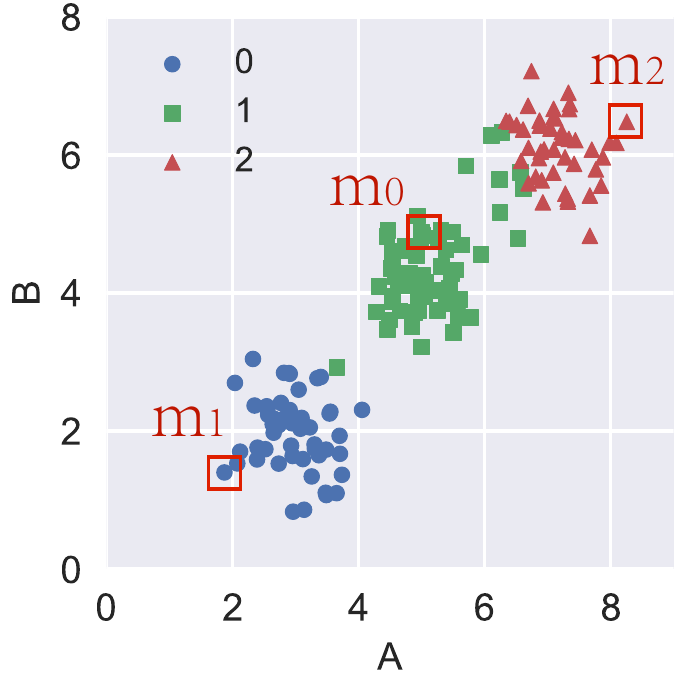}}
	\subfigure[Follow-up dataset]{ \label{subfig:MR5.1FFfollow}
		\includegraphics[width=0.23\textwidth]{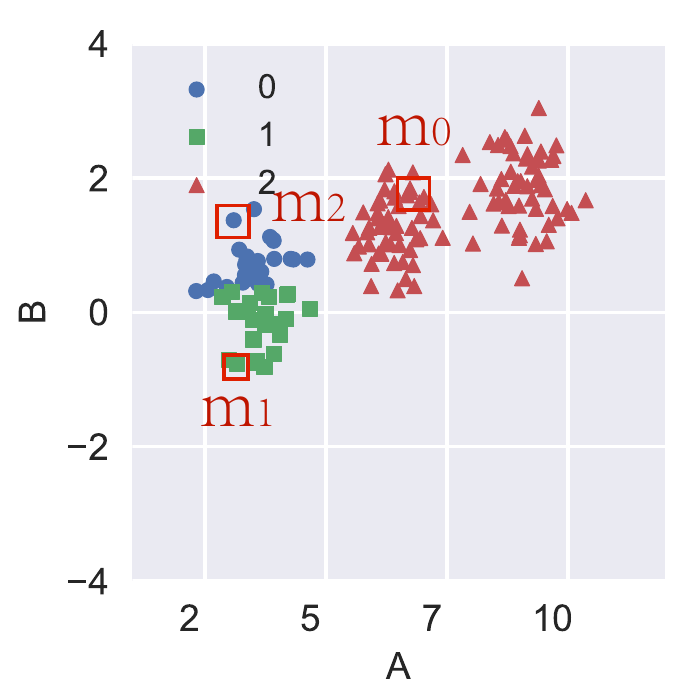}}
	\caption{Pattern type MERGE \& SPLIT for FF\@. The points enclosed in red boxes are the cluster centroids. $m_0$, $m_1$, and $m_2$ denote the first, second, and third selected centroids, respectively.}
	\label{fig:MR5.1FFSample}
\end{figure}

Similar to MR1.1, adding sample objects (MR3.1 and MR3.2) or inserting outliers (MR6) may change the farthest-first traversal sequence, thereby affecting the clustering results. For MR4.1, follow-up clusters should have better separation after adding informative attributes. However, unexpected results were still observed for FF.

For MR2.2, reclustering also occurred for the ``marginal'' points.
Moreover, we found that the source execution generated inaccurate results where the points on the margin of one cluster were assigned to another cluster, while the follow-up execution generated four well-clustered results.\,%
This observation revealed a reclustering problem of FF with respect to MR2.2\@.
As for MR5.1, we found that the violations were mainly due to the data normalization task during the preprocessing stage, and the effects of normalization varied across different violations.

For BORDER, only data samples near the boundaries were affected. For MERGE \& SPLIT, data normalization had a greater impact on the clustering result, which will be discussed in detail below.
	
\Nskip
\noindent
\textbf{MERGE \& SPLIT\@.}\,
	This pattern type was observed in $15\%$ violations to MR5.1 (rotating the coordinate system), with the $\RP$ values varied from $30\%$ to $50\%$. We noted from the Weka implementation that min-max normalization will be applied before computing the Euclidean distance of a pair of data objects. FF will randomly select a starting centroid $m_0$ as the first cluster centroid, and will then select a farthest point $m_1$ from $m_0$ as the second centroid (the remaining centroids will be selected in the same way). Eventually, every data point will be assigned to its nearest centroid. By rotating the coordinates, data assignment could be different due to the 
slight change on the normalized distance.
	
	Fig.~\ref{fig:MR5.1FFSample} illustrates how the traversal sequence is affected in relation to MR5.1\@. We obtained the "same" (i.e., the instances with the same index) starting centroid $m_0$ in the source and follow-up executions by fixing the random seed during the experiment. After FF had finished the first traversal, different points were chosen as the second centroids $m_1$ in the source and follow-up executions. Similarly, after completing the second traversal, the third centroids $m_2$ in the source and follow-up executions were different. In the end, the resulting clusters turned out to be totally different between the source and follow-up datasets. 

\Nskip
\begin{tcolorbox}[colframe=black!75!white, colback=white, boxrule=0.3mm]
	\emph{\textbf{Summary:}}\, The traversal sequence of FF largely depends on the starting centroid. After a data object has been assigned to a cluster, it can no longer be moved around. Therefore, FF is much more sensitive to data transformation such as reordering the data sequence and inserting outliers (or noises). We found that FF is effective in recognizing an outlier and assigning it to a single cluster, without being much affected by data transformation. However, data transformation may cause data objects other than outliers to be reassigned to different clusters. Furthermore, FF occasionally does not generate clearcut and accurate clusters as expected, even when the data samples are well separated.
\end{tcolorbox}

\Nskip

\Sskip
\subsubsection{Violations Related to DS}
\label{subsec:dbscan}

\begin{figure}[t]
	\centering
	\subfigure[Source dataset]{ \label{subfig:MR1DBsource}
		\includegraphics[width=0.22\textwidth]{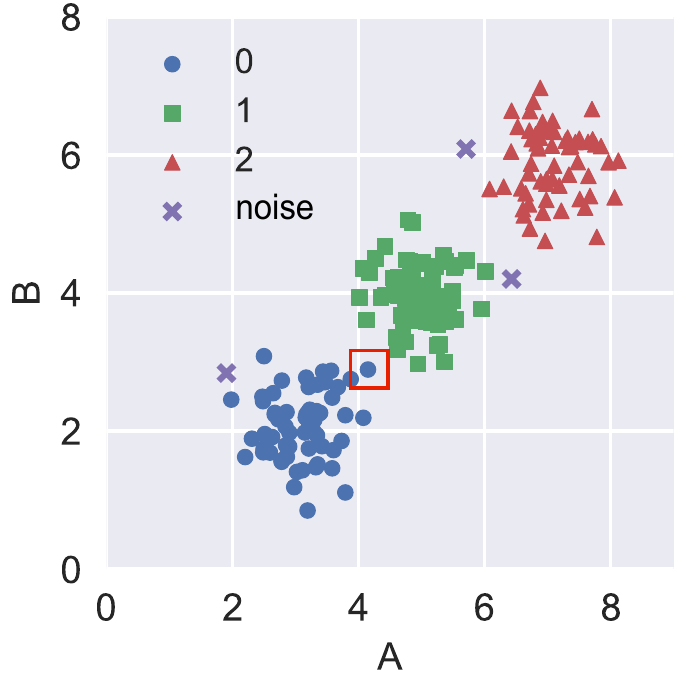}}
	\subfigure[Follow-up dataset]{ \label{subfig:MR1DBfollow}
		\includegraphics[width=0.22\textwidth]{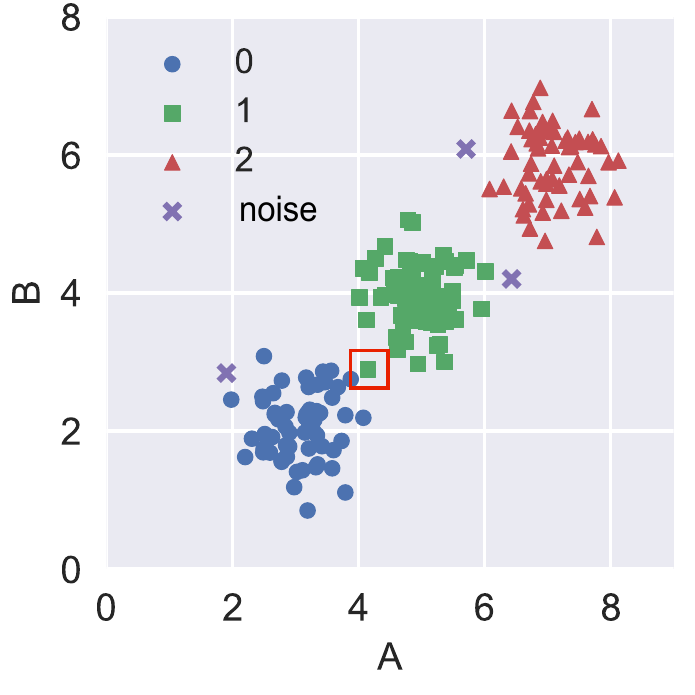}}
	\caption{Pattern type BORDER for DS. } \label{fig:MR1DBSCANSample}
\end{figure}

\begin{figure}[t]
	\centering
	\subfigure[Source dataset]{\label{subfig:MR2.2DBsource2}
		\includegraphics[width=0.22\textwidth]{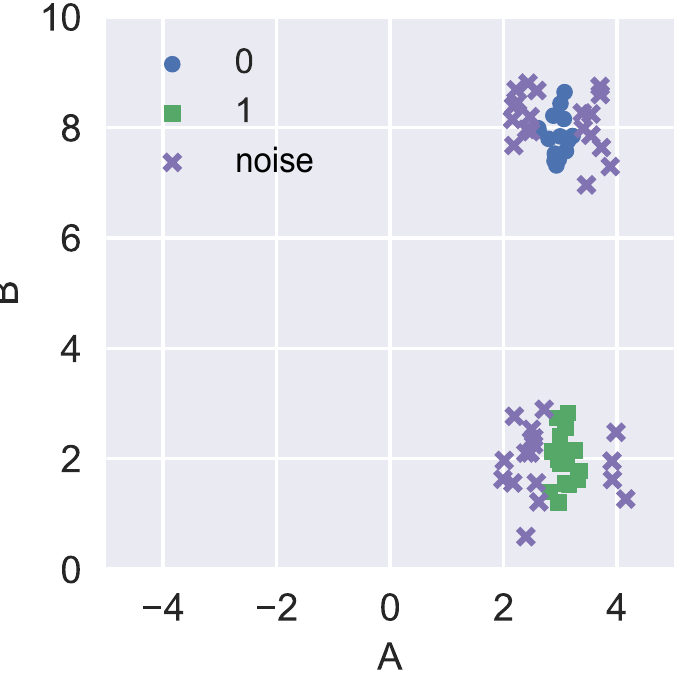}}
	\subfigure[Follow-up dataset]{\label{subfig:MR2.2DBfollow2}
		\includegraphics[width=0.22\textwidth]{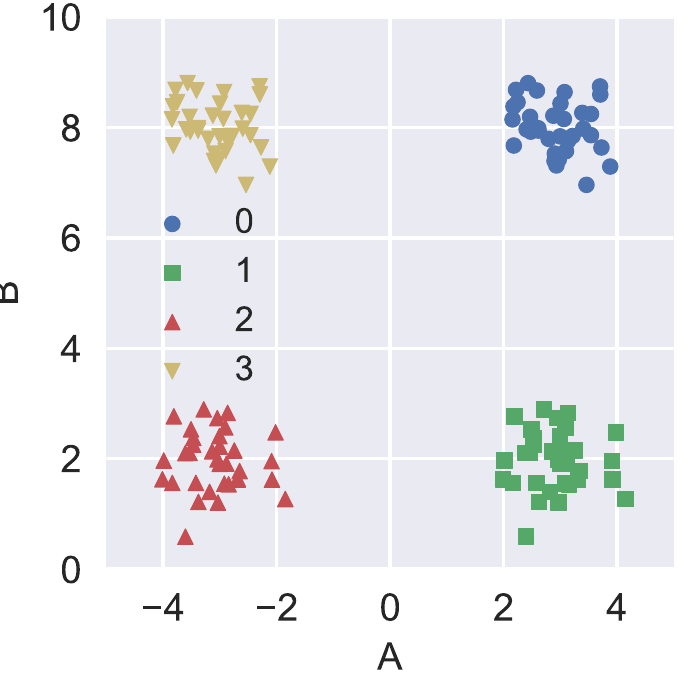}}
	\caption{Pattern type NOISE for DS\@. Blue dots in (a) denote the ''noisy'' data detected by DS, but there was no ''noisy'' data detected in (b).} \label{fig:MR2.2DBSample}
\end{figure}

	\begin{figure}[t]
	\centering
	\subfigure[Source dataset]{ \label{subFig:MR5.1DBsource1}
		\includegraphics[width=0.22\textwidth]{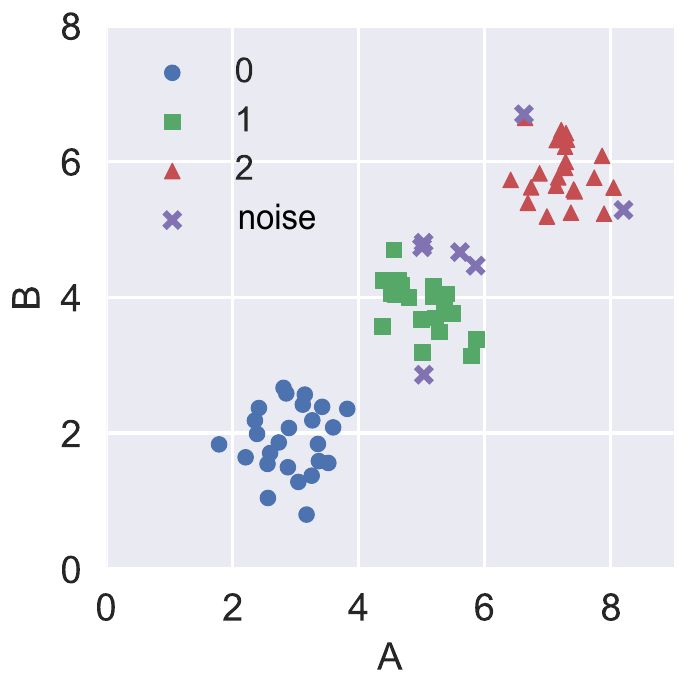}}
	\subfigure[Follow-up dataset]{ \label{subFig:MR5.1DBfollow1}
		\includegraphics[width=0.22\textwidth]{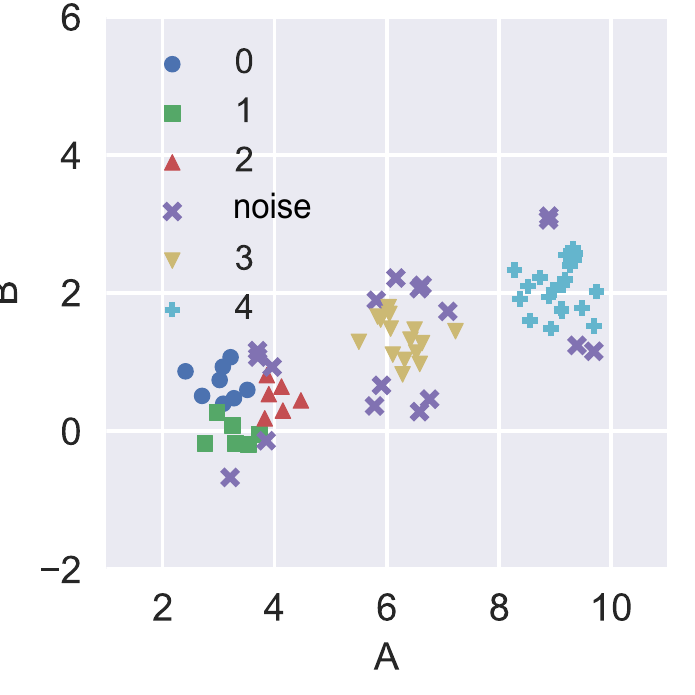}}
	\caption{Pattern type NUM of DS\@.}
	\label{fig:MR5.1DBSample}
\end{figure}

Fig.~\ref{subfig:DBRPDistribution} shows that violations to MR1.1, MR2.2, MR3.1, MR3.2, MR5,1 and MR6 occurred, with a wide range of $\RP$ values (between 0\% and 70\%). With further analysis, we noted that some points were ``noises'', representing a major difference on the clustering results between DS and other methods.

\Sskip
\noindent
\textbf{BORDER\@.}\, 
This pattern type was observed in the violations to MR1.1, with $\RP < 3\%$. In this pattern type, violations occurred near the cluster boundaries 
(see the \zzy{points in the two boxes} in Fig.~\ref{fig:MR1DBSCANSample}),
especially in those cases where clusters were close to each other.
It has been reported by others (e.g., in~\cite{ester1996density}) that DS is almost independent of the order of the input data objects. In our experiment, however, we observed that, among the 100 trials with MR1.1, eight violations related to BORDER occurred. In each of these violations, a very small portion of data objects was found to be assigned to different clusters from the source dataset to the follow-up dataset. These violations occurred due to a property of DS: if a data object was density-reachable from two neighbor clusters, the cluster to which this data object would be assigned was decided by the chronological sequence of detecting the clusters near that object.
Nevertheless, DS was fairly robust to the type of data transformation corresponding to MR1.1 if data samples to be clustered were well separated.

\Sskip
\noindent
\textbf{NOISE\@.}\,
For DS, we observed another violation pattern type related to noisy data.
This pattern type occurred in all the violations to MR2.2, MR3.1, MR4.2, and MR6.1; in 86\% (18 out of 21) violations to MR3.2; and in 90\% (55 out of 61) violations to MR5.1\@.
	
Consider MR2.2 (data mirroring) as an example. We noted from Fig.~\ref{fig:MR2.2DBSample} that some points marked as ``noisy'' data in the source clusters turned out to be density-reachable points in the follow-up clusters. We also noted that the number of noisy data sharply dropped to zero or a tiny value after performing data mirroring as prescribed by MR2.2\@. We also observed that DS only generated good clustering results when the parameters $\mathit{eps}$ and $\mathit{minPts}$ were properly set. More specifically, when these two parameters were properly set so that noisy data did not occur in the source dataset, then no violation to MR2.2 would have occurred in the follow-up dataset.
Similarly, for those violations to MR3.1, MR3.2, MR4.2, and MR6.1, the number of "noises'' also decreased after performing the types of data transformation corresponding to these MRs.

By analyzing the implementation of DS, the above violations can be explained as follows. Suppose $q$ denotes a density-connected point from a core point $p$ in cluster $c$; $o$ denotes a point in the $\mathit{eps}$-neighborhood of $q$ which is marked as ``noise''. After inserting new data points to $c$, there may be as many points as $\mathit{minPts}$ within $q$'s $\mathit{eps}$-neighborhood. Thus, $q$ becomes a new core point, and $o$ is density-connected to $p$ so that $o$ becomes a new member of cluster $c$. Hence, the number of noises is expected to decrease or remain unchanged after inserting new data points to a cluster. Our analysis result can be seen as a convenient quantification of the execution behavior of DS\@.

\Sskip
\noindent
\textbf{NUM\@.}\,
This pattern type was observed in three violations to MR3.2, and in six violations to MR5.1, where DS recognized an incorrect number of clusters. 
	Take MR5.1 as an example. We found that DS unexpectedly divided the data samples into four or more clusters, 
 and at the same time the number of samples labeled as "noises'' 
 \zzy{(see the data points ``$\times$''}
in Figs.~\ref{subFig:MR5.1DBsource1} and~\ref{subFig:MR5.1DBfollow1}) increased from the source dataset to the follow-up dataset. Since DS in the Weka implementation includes an embedded data normalization routine, the generated clustering result could be affected by even a slight change on the normalized distance among data objects.

\Sskip
\begin{tcolorbox}[colframe=black!75!white, colback=white, boxrule=0.3mm]
	\emph{\textbf{Summary:}}\, Although the clustering result of DS is generally considered as not being affected by the input order of data samples, our experiment revealed that this was not the case due to the randomness of the system itself. When compared with other clustering systems, DS is effective in recognizing outliers.
	We also found that ``nosiy" data points are sensitive to data transformation. The configuration of parameters which may have some impacts on the clustering result is also important. As a whole, DS is robust to different types of data transformation\,---\,this is what users expect on a clustering system. 
\end{tcolorbox}

\subsection{Summary and Further Analysis}
\label{subsec:summary}

We learnt from the analyses and discussions in Sections~\ref{subsec:km} to~\ref{subsec:dbscan} that, for all subject clustering systems, data samples located near the cluster boundaries were sensitive to even a small change in data input. This can be explained by the randomness of the system during its initialization. 
Moreover, those systems (KM, XM, and FF) which largely depend on the initialization conditions showed a larger impact of data transformation on the clustering result, while EM, AN and DS showed higher robustness to such change.
Undoubtedly, users normally expect that the chosen clustering system will have high robustness to relocating data samples from near the boundary of one cluster to another cluster as a result of data transformation. Thus, in this aspect, EM, AN and DS are more preferable than the other systems.

\ppl{
	Table~\ref{tb:summary} summarizes, for each subject clustering system, its compliance with or violations to the relevant generic MRs. Furthermore, for each violation case, we give the plausible reason(s) for its occurrence. Consider, for example, the cell related to KM and MR1.1. This cell indicates that KM exhibited two types of violation patterns (BORDER and MERGE\,\&\,SPLIT) with respect to MR1.1. The cause of BORDER was due to the random initialization of the cluster centroid. For MERGE\,\&\,SPLIT, it occurred because KM was trapped into a local optimum. Table~\ref{tb:summary} not only summarizes our assessment results of the subject clustering systems, but also serves as a useful and handy checklist for users to make informed decisions on which clustering systems to choose with respect to their expectations.
	Also, users are allowed to assign different weights to different violation patterns. For example, if users consider that violations with the BORDER pattern are less important than violations with the MERGE\,\&\,SPLIT pattern, then MERGE\,\&\,SPLIT can be assigned a higher weight than BORDER\@. In this way,
	the compliance with each generic MR (in Table~\ref{tb:summary}), together with its corresponding weighted score, can be used as a test adequacy criterion to be further leveraged by users for selecting an appropriate clustering system in accordance with the users' expectations.
In addition, we also analyzed and summarized the relative strengths and weaknesses of the six subject systems with respect to the 11 generic MRs in Table~\ref{tb:all}, with a view to facilitating users to gain a deep understanding of these systems.}

\afterpage{
	\begin{landscape}
		\begin{table}[t]
			\renewcommand\arraystretch{3}
			\caption{\ppl{Summary of Compliance and Violation Patterns (with Plausible Reasons) with Respect to Each Generic MR}}\label{tb:summary}
			\centering
			\scalebox{0.85}{
				\begin{tabular}{lp{4.5cm}<{\raggedright}p{4cm}<{\raggedright}p{4cm}p{4cm}
						                   p{4cm}p{4.2cm}<{\raggedright}}
					\toprule
					MR & \multicolumn{6}{l}{Clustering Systems} \\
					\cmidrule{2-7}
					& KM & XM & EM & AN & FF & DS \\
					\midrule
					MR1.1 & \emph{BORDER}: random initialization of (starting) centroids; \emph{MERGER\,\&\,SPLIT}: tends to converge to local optima
					& \emph{BORDER}: random initialization of (starting) centroids; \emph{MERGER\,\&\,SPLIT}: tends to converge to local optima
					& \checkmark
					& \checkmark
					& \emph{BORDER}: random initialization of the starting centroid
					& \emph{BORDER}: chronological sequence of detecting clusters \\
					MR1.2 & \checkmark & \checkmark & N/A & N/A & \checkmark & N/A \\
					MR2.1 & \emph{MERGE\,\&\,SPLIT}: tends to converge to local optima
					& \checkmark & \checkmark & N/A & \checkmark & N/A \\
					MR2.2 & \emph{MERGE\,\&\,SPLIT}: tends to converge to local optima
					& \checkmark & \checkmark & \checkmark
					& \emph{BORDER}: change of traversal sequence
					&  \emph{NOISE}: noisy objects become density-reachable objects  \\
					MR3.1 & \emph{BORDER}: random initialization of (starting) centroids; \emph{MERGER\,\&\,SPLIT}: tends to converge to local optima
					& \emph{BORDER}: random initialization of (starting) centroids; \emph{MERGER\,\&\,SPLIT}: tends to converge to local optima
					& \emph{BORDER}: deviation in parameter estimation
					& \emph{BORDER}: deviation in distance computation
					& \emph{BORDER}:  random initialization of the starting centroid
					& \emph{NOISE}: decrease in noisy objects (increase in core objects)  \\
					MR3.2 & \emph{BORDER}: random initialization of (starting) centroids; \emph{MERGER\,\&\,SPLIT}: tends to converge to local optima
					& \emph{BORDER}: random initialization of (starting) centroids; \emph{MERGER\,\&\,SPLIT}: tends to converge to local optima
					& \emph{BORDER}: deviation in parameter estimation
					& \emph{BORDER}: deviation in distance computation
					& \emph{BORDER}:  random initialization of the starting centroid
					& \emph{NOISE}: increase in density-reachable points  \\
					MR4.1 & \emph{MERGE\,\&\,SPLIT}: tends to converge to local optima
					& \checkmark  & \checkmark  & \checkmark 
					& \emph{BORDER}:  change of traversal sequence
					& \checkmark   \\
					MR4.2 & \emph{MERGE\,\&\,SPLIT}: tends to converge to local optima
					& \checkmark  & \checkmark  & \checkmark 
					& \checkmark 
					&  \emph{NOISE}:   increase in density-reachable points \\	
					MR5.1 & 	\emph{BORDER}: min-max normalization; \emph{MERGER\,\&\,SPLIT}: tends to converge to local optima
					& 	\emph{BORDER}: min-max normalization; \emph{MERGER\,\&\,SPLIT}: tends to converge to local optima
					&  \emph{BORDER}: min-max normalization; \emph{SPLIT}: suboptimal KM solution during initialization          
					& \emph{BORDER}: min-max normalization; \emph{SPLIT}: min-max normalization       
					& \emph{BORDER}: min-max normalization; \emph{MERGE\,\&\,SPLIT}: min-max normalization       
					&   \emph{BORDER}: min-max normalization; \emph{NUM}: min-max normalization ;      \\
					MR5.2 & \checkmark  & \checkmark  & \checkmark  & \checkmark  & \checkmark  & \checkmark  \\
					MR6 & \emph{BORDER}: random initialization of (starting) centroids; \emph{MERGER\,\&\,SPLIT}: tends to converge to local optima
					& \emph{BORDER}: random initialization of (starting) centroids; \emph{MERGER\,\&\,SPLIT}: tends to converge to local optima
					& \emph{BORDER}: deviation in parameter estimation
					& \checkmark 
					& \emph{BORDER}: deviation in distance computation
					& \emph{NOISE}: decrease in noisy objects \\			           
					\bottomrule
			\end{tabular}}	
			\vspace{0.15cm}
			{\footnotesize
				\begin{flushleft} \dag~~%
					``\checkmark'' means that no violation has been revealed by the relevant MR, and ``N/A'' means that the relevant MR is inapplicable to the clustering system.
				\end{flushleft}
			}
		\end{table}
	\end{landscape}
}

\ppl{Surprisingly, our experimental results (see Table~\ref{tb:VR} and~\ref{tb:RP}) show that all the subject systems involved many violations to MR5.1 (rotating the coordinate system).}
A close examination of the corresponding source code found that min-max normalization is the major cause of the observed violations. More specifically, the normalized distance among data points could be different after nonlinear data transformation such as rotating the coordinates (even if the data distribution remains unchanged). 
Note that data normalization is a very important step in most machine learning systems\,---\,some of these systems (e.g., those available in Weka) have embedded a data normalization routine in them. 
\zzy{Without using {\sc mettle}, users are unlikely to get an opportunity to understand the impact of the embedded normalization routine in machine learning systems.}

\ppl{The above discussion shows that, apart from assessing clustering systems and facilitating their selection, {\sc mettle} also supports program comprehension and end-user software engineering~\cite{Segal2005, Burnett2009}, through which users can gain a deeper understanding of the program under test without the need for using relevant complex theories.}

\afterpage{ 
\begin{table*}[t]
	\renewcommand\arraystretch{1.3}
	\caption{Summary of Relative Strengths and Weaknesses of Six Subject Clustering Systems}\label{tb:all}
	\centering
	\scalebox{1}{
		\begin{tabular}{lp{7.5cm}p{7.5cm}} \toprule
			Clustering & & \\
			Systems & Strengths & Weaknesses \\
			\midrule
			KM  &  Easy to understand and use & Sensitive to the random initialization of centroids and sensitive to outliers; may easily be trapped into a local optimum which may in turn lead to the occurrence of the MERGE\,\&\,SPLIT pattern   \\
			XM & A partial remedy for the local optimum problem; shows great advantages over KM when data groups are sufficiently separated &  Occasionally tends to underestimate the true number of clusters given a range $[k_{min}, k_{max}]$  \\
			EM & Strongly robust to various types of data transformation; less sensitive to a local optimum than KM & Depends partially on the KM solution during the initialization stage (EM initializes estimators by running KM $10$ times and then choosing the "best" solution with the smallest squared error) and, hence, the clustering result is sometimes not what the users expect \\
			AN & Independent of the order of input data; fairly robust to data transformation; strong ability to recognize outliers & Scaling datasets will possibly change the clustering results   \\	
			FF &  Involves less reassignment and adjustment, which may speed up the clustering process & Very sensitive to initialization conditions; sensitive to even a small change in data input; occasionally generates inaccurate clusters even when the data samples are well separated   \\
			DS &  
			Effective in recognizing outliers
			& The values of the parameters $\mathit{eps}$ and $\mathit{minPts}$ would have a large impact on the clustering result; noisy data are often relocated to different clusters after data transformation    \\
			\bottomrule
	\end{tabular}}
	\vspace{0.1cm}
\end{table*}
}

\section{mettle as a framework for selecting clustering systems}
\label{sec:application}

Apart from assessing clustering systems, another potential application of {\sc mettle} is to help users select the most appropriate clustering systems.
With more and more open-source software libraries that provide ready-to-use machine learning systems, users are facing a big challenge in choosing a proper one for their application scenarios. 
Traditionally, users apply a \emph{data-driven} approach to tackle this challenge, where a set of candidate systems are run against various datasets. After execution, cross-validation and statistical analyses are used to help users select the proper system to use~\cite{Caruana2006, Fern2014, rohtua2018}. 
However, we argue that, besides the average performance of a clustering system across various datasets, users' expectations or requirements on the system with respect to the application scenario should also be taken into account. 

Following this argument, {\sc mettle} does provide an intuitively appealing and systematic framework to aid selecting proper clustering systems, by enabling users to assess the appropriateness of these systems based on their own \emph{specific} requirements and expectations. Below we give more detailed explanation.

First, the framework of METTLE involves a concept of ``adequacy criterion''. For example, a list of generic MRs derived from users' expectations is used in {\sc mettle} as an adequacy criterion. Subject clustering systems are then assessed by validating the compliance with each generic MR. The results of assessment are used for selecting an appropriate system in accordance with users' own specific needs.

Test adequacy plays a crucial role in traditional software testing and validation. A lot of coverage criteria from different perspectives have already been proposed to measure test adequacy, such as statement coverage, branch coverage, and path coverage~\cite{zhu1997}.
The necessity for evaluating test adequacy has been gradually accepted in machine learning testing~\cite{mltestingsurvey}. Many researchers from the software engineering community have been working on proposing suitable criteria for evaluating the test adequacy for machine learning systems with a view to gaining confidence on the testing results~\cite{deepgauge,surprise}.
However, until now, there have been very few generally acceptable and systematic criteria for users to assess and validate machine learning (include clustering) systems in their own contexts. 

Traditional clustering assessment methods can be regarded as a type of data-oriented adequacy measurement, by exploring the ``adequacy'' in the input space. However, with such data-oriented adequacy criterion, users cannot easily link the input to the appropriateness of a system with respect to their own expectations and requirements. 
In contrast,  our {\sc mettle} provides a  property-oriented adequacy criterion based on MRs, which can easily address the above problem in traditional methods. In fact, this property-oriented adequacy criterion makes the first step in the potential research direction pointed out by \citeauthor{chen2018metamorphic} \cite{chen2018metamorphic}, where they argue that MT can allow the development of an MR-based metric to be used as a black-box test adequacy criterion.
Assessing the compliance with MRs provides useful information about the quality and appropriateness of the relevant properties and functionalities of a clustering system in a particular application domain. Thus, such an MR-based criterion in {\sc mettle} can provide more confidence to users in making decision about which clustering system to select.

Table~\ref{tb:summary} summarizes the performance in terms of the compliance with each generic MR of the six subject clustering systems with respect to the 11 generic MRs.  As discussed in Section~\ref{subsec:summary}, this table can be used to help users make informed decision about which clustering system to select for use in a specific scenario. In addition to adopting some or all of the 11 generic MRs, more specific MRs can be defined by users to complement the generic ones (if users have expectations that do not correspond to any of these generic MRs). Note that users are not required to have substantial and sophisticated knowledge on the candidate clustering systems. This is because defining specific MRs is primarily based on users' domain knowledge of their applications.
The adopted generic MRs, together with the additional, specific MRs defined by users, form a comprehensive checklist where MR compliance and the associated weighted scores can be used as a selection criterion.

In reality, a user may not consider all selected MRs (and their corresponding types of data transformation) to be equally important. In other words, some selected MRs are considered more preferable while the others are less preferable. Consider, for example, an e-commerce firm with a fast-growing number of online customers. Each of these customers has a registered account with the e-commerce firm. Consider further the following scenarios: 

\Sskip
{\bf Scenario~1.} The marketing department of the e-commerce firm often clusters its customers into different groups to facilitate new product recommendation to the targeted groups. In this case, the marketing director may be highly concerned with the impact of adding data samples (correspond to newly registered customer accounts) near a cluster's centroid or boundary on the clustering result generated by a clustering system.

\Sskip
 {\bf Scenario~2.} The business fraud department of the e-commerce firm may concern more on how a clustering system handles outliers because they may correspond to malicious hackers. 

\Sskip

In view of the different levels of importance on the types of data transformation (and their corresponding MRs), the overall framework to support users to select clustering systems (in the context of {\sc mettle}) is given as follows:

\Sskip
\begin{itemize}
	
	 \item[(1)] Select generic MRs or define new MRs in accordance with the user's intuitive expectations and specific requirements related to their application domains.
	
    \Sskip
	\item [(2)] Classify all the selected MRs into two categories: ``must have'' and ``nice to have''.
	
	\Sskip
	\item [(3)] Use {\sc mettle} to validate all the candidate clustering systems against all the selected MRs by executing each method twice (first with the source dataset, then with the follow-up dataset).
	
	\Sskip
	\item[(4)] Construct a summary table \ppl{which summarizes the violation patterns with respect to all the selected MRs.}
	
	\Sskip
	\item [(5)] \zzy{For each ``nice-to-have'' selection MR, assign a weight $w_1$ (where $0.0 < w_1 < 1.0$), so that a higher value of $w_1$ means that the corresponding MR is relatively more preferable or important. Then, assign a weight $w_2$ (where $0.0 < w_2 < 1.0$) according to the type of violation patterns related to this MR, so that a higher value of $w_2$ indicates more severity for the corresponding violation pattern.}
	
	\Sskip
	\item [(6)] Ignore those clustering systems which show violations to any ``must-have'' MR\@.
	
	\Sskip
	\item [(7)] For every remaining clustering system $m_i$ (where $1 \leq i \leq k$; $k$ = total number of remaining \ppl{systems}), calculate its score $S_{m_i}$ using the following formula:
	
	\begin{equation*}
	\begin{split}
		S _{m_i} = & (w_{11} \times w_{12} \times x_1) + (w_{12} \times w_{22} \times x_2) + \\
	                    & \cdots + (w_{1n} \times w_{2n} \times x_n)
	\end{split}
	\end{equation*}
	
	\Sskip
	where $1 \leq j \leq n$; $n$ = total number of selection MRs; $w_j$ = the weight assigned to MR$_j$; \zzy{$x_j = 1$ if one or more violations to MR$_j$ occur, $x_j = 0$ if no violation to MR$_j$ occurs.}
    
    \Sskip
	\item [\ppl{(8)}] The most appropriate system to select is the $m_i$ with the \zzy{\emph{smallest}} $S_{m_i}$.
\end{itemize}

%



\Nskip
\noindent
By means of the above selection framework, \ppl{users are able to devise their own quality assessment schemes} for evaluating a set of candidate clustering systems in accordance with their own preferences.

As a reminder, the individual lists of selected MRs developed by different users in the same application domain can be shared, with a view to developing a more comprehensive and effective aggregated list of selection MRs. Furthermore, a repository (e.g., in~\cite{METWiki}) can be created to store all the selected MRs and their corresponding validation results for some clustering systems. Via this repository, even inexperienced users without much knowledge about the execution behaviors of individual clustering systems (with respect to different types of data transformation) can still effectively evaluate and then select their most preferred systems.

\section{Threats to Validity}
\label{sec:threats}

In this section, we discuss \ppl{some potential factors} that might affect the validity of \ppl{our experiment.}

\Sskip
\subsection{Internal Validity}
\label{subsec:intenal validity}

A main internal threat to our study is the randomness of clustering systems. Some of these systems will randomly select an object from the dataset in their initialization. This may lead to result variations across multiple system executions. To alleviate this threat, we fixed the random seed for the relevant systems in our experiment, so that the clustering results are reproducible in each execution run.

Another internal threat is related to parameter configuration. Different input parameters would lead to totally different clustering results. Thus, the impact of parameters on clustering validity is definitely a further research topic for clustering assessment and validation. For example, DS has two critical parameters: \ppl{the} minimal number of points $\mathit{minPts}$ within a specific distance $\mathit{eps}$. The \ppl{clustering result generated by DS is largely affected by} these two parameters. In this paper, we do not \ppl{attempt} to evaluate the impact of different parameters on the clustering result. Thus, these parameters were not treated as independent variables and, hence, were fixed during our experiment.

\subsection{External Validity}
\label{subsection:external validity}

In {\sc mettle}, we leveraged the concept of MT and developed a list of generic MRs for validation. Because these generic MRs do not cover all possible properties of clustering systems, this issue is therefore a potential threat to the external validity of our study. However, as a novel assessment and validation framework based on the users' perspective, {\sc mettle} allows users to specify their expected characteristics of a clustering system in their own contexts. In {\sc mettle}, users could simply adopt some of all of the 11 generic MRs we developed, and then supplemented by more specific, user-defined MRs according to their own application scenarios.

As an application of MT, {\sc mettle} also has limitations in some areas of cluster analysis, for example, identifying the optimal number of clusters. MT was proposed to alleviate (not to completely solve) the oracle problem in software testing. Also, by its very nature, an MR can only reveal the absence of an expected characterization from the system, rather than computing the correctness of individual outputs.

Another external threat is the generality of our approach. In this regard, it is well known that in the field of cluster validation, there does not exist a single validation approach which can effectively handle all dataset types~\cite{pal1995}. {\sc mettle} is no exception. Our experiment only involved those datasets with well-formed clusters so that all the six subject clustering systems could properly handle these clusters. A similar approach to generating synthetic datasets for experiments has also been adopted in some other studies (e.g., \cite{huang2001}). Although the datasets used for assessment may vary case by case, the high-level properties of clustering systems to be assessed and evaluated by {\sc mettle} are rather general. Thus, we argue that the effect of different datasets on the effectiveness of {\sc mettle} should not be large.

\section{Related Work}
\label{sec:related}

MT has been successfully applied in many applications since its introduction by \citeauthor{chen1998metamorphic}~ \cite{chen1998metamorphic}. We refer the readers to recent surveys on MT~\cite{chen2018metamorphic,7422146} to gain further insight into this technique. In this section, we highlight some recent work on MT by both academia and industry researchers.

\citeauthor{zhou2016metamorphic} \cite{zhou2016metamorphic} proposed a user-oriented testing approach for the quality assessment of major online search engines (including, for example, Google and Bing) using the concept of MT\@. Their empirical results not only guide developers to identify the weaknesses of these search engines, but also help users choose a proper online search engine in a specific scenario. 
\citeauthor{8074764} \cite{8074764} applied MT to web application programming interfaces (APIs) for automatic fault detection. They first constructed MRs with an output-driven approach, and then applied their method to APIs of Spotify and YouTube. This application successfully detected 11 real-life problems, indicating the effectiveness of MT\@.
A recent work~\cite{Dwarakanath:2018:IIB:3213846.3213858} has been reported, which is related to using MT for software verification of machine-learning-based image classifiers. The effectiveness of MRs was tested by mutation testing, where 71\% implementation faults were successfully caught.

Adding to the successful applications of MT to quality assessment as well as software verification and validation, MT has also been applied to detecting performance bugs~\cite{SEGURA20181}. In this work, a set of performance MRs was defined for the automatic analysis of feature models. A proof-of-concept experiment was conducted to confirm the feasibility of using a metamorphic approach to detecting performance faults.

In recent years, we have witnessed the advances in deep learning. Applying MT to AI-driven systems has grown rapidly. 
In~\cite{7961649}, MT was used to validate the classification accuracy of deep learning frameworks. Also, DeepTest, a testing tool for Deep-Neural-Network-driven autonomous vehicles, was developed to leverage MRs to create a test oracle~\cite{Tian:2018:DAT:3180155.3180220}. DeepTest automatically generates synthetic test cases for different real-world conditions, and is able to detect thousands of erroneous behaviors in autonomous driving systems. Furthermore, a framework called DeepRoad \cite{Zhang:2018:DGM:3238147.3238187} was proposed for testing autonomous driving system, with a view to detecting inconsistent behaviors across various synthesized driving scenes based on MRs.

More recently, an internationally renowned IT consultancy and service firm, Accenture, has applied MT to test machine learning systems, providing a new vision for quality engineering~\cite{Accenture}. 
In addition, GraphicsFuzz, a commercial spin-off firm from the Department of Computing at Imperial College London, has pioneered the combination of fuzzing and MT for testing graphics drivers~\cite{GraphicsFuzz}. 
GraphicsFuzz toolset has been successful at exploring defects in a large number of graphics driver across different platforms, for example, an Shield TV box with an NVIDIA GPU and Samsung Galaxy S9 with an ARM GPU.
GraphicsFuzz was later acquired by Google in August 2018~\cite{GoogleAcquires}.

\section{Conclusion and Future Work}
\label{sec:conclusion}

In this paper, we propose a metamorphic testing-based approach ({\sc mettle}) to assessing and validating clustering systems by considering the various dynamic data perspectives for different application scenarios. 
We have defined $11$ generic metamorphic relations (MRs) for six common types of data transformation. We have used these generic MRs, together with six subject clustering systems, to conduct an experiment for verifying the viability and effectiveness of {\sc mettle}. Our experiment has demonstrated that {\sc mettle} is a vivid, flexible, and practical approach towards validating and assessing clustering systems.

In general, {\sc mettle} has the following merits with respect to validation and assessment:

\begin{itemize}
       \item {\bf Validation}

\begin{itemize}
	\item It is generic and can be easily applied to any clustering systems.

\Sskip	
	\item It provides an elegant and tailor-made mechanism for end users to define their specific expectations and requirements (in terms of MRs) when validating clustering systems.
	
\Sskip
	\item It is further supported by a set of $11$ generic MRs, which can be mostly applied to various clustering scenarios.
\end{itemize}

\item {\bf Assessment}

\begin{itemize}		
        \item It provides an innovative approach to unveil the characteristics of unsupervised machine learning systems.

\Sskip	
	\item It helps categorize clustering systems in terms of their strengths and weaknesses with respect to a set of MRs (corresponding to different types of data transformation). This is particularly helpful for those end users who are not knowledgeable about the \zzy{logic and mechanisms} of clustering systems.
	
\Sskip	
	\item It allows end users to devise their own quality assessment schemes for evaluating a set of candidate clustering systems (with respect to the user-defined MRs and their corresponding weights).  
	
	\item \zzy{It demonstrates a \ppl{systematic and practical} framework for end users to assess and  select appropriate clustering system for use.}
\end{itemize}

\end{itemize}

\Nskip
The promising and encouraging work described in this paper can be extended into \zzy{three} aspects.
First, it would be worthwhile to conduct another experiment involving high-dimensional data samples (the experiment described in this paper only involved datasets in two-dimensional space for easy visualization of the clustering results).
Secondly, it would be fruitful to investigate the issue of how to define good and representative MRs (in addition to the 11 generic ones) that are applicable to a wide range of application scenarios.
\zzy{\ppl{Thirdly}, the correlation between a violation to an MR and a particular error pattern certainly represents an interesting research topic that warrants further investigation.}

\section*{Acknowledgment}
This work was supported by the National Key R\&D Program of China under the grant number 2018YFB1003901, and the National Natural Science Foundation of China under the grant numbers 61572375, 61832009, and 61772263.

\ifCLASSOPTIONcaptionsoff
\newpage
\fi

\bibliographystyle{IEEEtranN}
\bibliography{Notes}

\end{document}